\documentclass[draftcls,onecolumn,12pt]{IEEEtran}

\usepackage{graphicx}
\usepackage{amsmath,amsthm}
\usepackage{amsfonts}
\usepackage{booktabs}
\usepackage{tabularx}
\usepackage{subfigure}
\usepackage{setspace}

\ifCLASSOPTIONcompsoc
\else
\fi
\ifCLASSINFOpdf
\else
\fi
\begin{document}
%
\title{Hierarchic Power Allocation for Spectrum Sharing in OFDM-Based Cognitive Radio Networks}

\author{Tian~Zhang,
Wei~Chen,~\IEEEmembership{Member,~IEEE,}
        Zhu Han,~\IEEEmembership{Senior Member,~IEEE,}
        and~Zhigang~Cao,~\IEEEmembership{Senior Member,~IEEE}
        \thanks{This work is partially supported by the National
Basic Research Program of China (973 Program) under
Grant 2012CB316001, the National Nature Science
Foundation of China (NSFC) under Grants 60832008 and 60902001, US NSF CNS-1117560, CNS-0953377,
ECCS-1028782, CNS-0905556, and Qatar National Research Fund.}
\thanks{ T. Zhang is with the School of Information Science and Engineering, Shandong University, Jinan 250100, China. He is also with the State Key Laboratory on Microwave and Digital Communications, Department of Electronic Engineering,
Tsinghua National Laboratory for Information Science and Technology (TNList), Tsinghua University.
E-mail: tianzhang.ee@gmail.com}
\thanks{ W. Chen and Z. Cao are with the State Key Laboratory on
Microwave and Digital Communications, Department of Electronic Engineering,
Tsinghua National Laboratory for Information Science and Technology
(TNList), Tsinghua University, Beijing 100084, China.
E-mail: \{wchen, czg-dee\}@tsinghua.edu.cn}
\thanks{ Z. Han is with the Department of Electrical and Computer Engineering, University
of Houston, Houston, TX 77004, USA.
E-mail: zhan2@mail.uh.edu}
}

%


\maketitle

\begin{abstract}
In this paper, a Stackelberg game is built to model the hierarchic power allocation of primary user (PU) network and secondary user (SU) network in OFDM-based cognitive radio (CR) networks. We formulate the PU and the SUs as the leader and the followers, respectively. We consider two constraints: the total power constraint and the interference-to-signal ratio (ISR) constraint, in which the ratio between the accumulated interference and the received signal power at each PU should not exceed certain threshold. Firstly, we focus on the single-PU and multi-SU scenario. Based on the analysis of the Stackelberg Equilibrium (SE) for the proposed Stackelberg game, an analytical hierarchic power allocation method is proposed when the PU can acquire the additional information to anticipate SUs' reaction. The analytical algorithm has two steps: 1) The PU optimizes its power allocation with considering the reaction of SUs to its action. In the power optimization of the PU, there is a sub-game for power allocation of SUs given fixed transmit power of the PU. The existence and uniqueness for the Nash Equilibrium (NE) of the sub-game are investigated. We also propose an iterative algorithm to obtain the NE, and derive the closed-form solutions of NE for the perfectly symmetric channel. 2) The SUs allocate the power according to the NE of the sub-game given PU's optimal power allocation. Furthermore, we design two distributed iterative algorithms for the general channel even when private information of the SUs is unavailable at the PU. The first iterative algorithm has a guaranteed convergence performance, and the second iterative algorithm employs asynchronous power update to improve time efficiency. Finally, we extend to the multi-PU and multi-SU scenario, and a distributed iterative algorithm is presented.
\end{abstract}
%
%
%


%
\IEEEpeerreviewmaketitle

\section{Introduction}
%
%
%
%

%

\IEEEPARstart{C}{ognitive} radio (CR) technology has gained much attention because of its capability of improving the spectrum utilization efficiency \cite{PC99:J. Mitola G. Q. Maguire}.
In CR networks, the CRs transmit in an opportunistic way or coexist with the primary systems simultaneously under the constraints that
the primary systems will not be harmed.  \par
Due to scarcity of power and hostile characteristics of wireless
channels, efficient power allocation schemes are necessary for design of high-performance CR networks.
Meanwhile, as the game theory is suitable for analyzing conflict and cooperation
among rational decision makers, it has emerged as a very powerful
tool for power allocation in CR networks \cite{JC09:J. W. Huang and V. Krishnamurthy,Book:Z.Han}. In the game theory based power allocation frameworks, the nodes are modeled as self-interested or group-rational players, and compete or cooperate with each other to maximize their utilities by viewing the power as the strategies.
The cooperative game theoretic approach of optimal power control for secondary users (SUs) in CR networks has been proposed in \cite{TVT10:C.-G. Yang J.-D. Li and Z. Tian}; the authors transformed the coupled interference constraints into a pricing function in the objective utility, and then the Kalai-Smorodinsky (KS) bargaining solution and the Nash bargaining solution (NBS) of the reformulated game were investigated. In \cite{SECON05:L. Cao and H. Zheng}, a fair local bargaining
framework was proposed for spectrum allocation, and two bargaining strategies named as one-to-one fairness
bargaining and feed poverty bargaining were presented.
The opportunistic
spectrum access problem was addressed by utilizing the cooperative game theory
in \cite{TWC09:J. Suris L. A. DaSilva Z. Han A. B. MacKenzie and R. S. Komali}, three bargaining solutions were compared
and analyzed, and a distributed algorithm that can achieve the NBS for the spectrum sharing game was presented.
In \cite{ICC10:O. N. Gharehshiran A. Attar and V. Krishnamurthy}, the authors investigated the resource
allocation in CR networks by using the coalitional game theory, and a distributed dynamic coalition formation algorithm was proposed.
A distributed power control protocol for the secondary network based on non-cooperative game was studied in \cite{ICC10:Y.-E. Lin. K.-H. Liu. and H.-Y. Hsieh}. Utilizing the best response, a distributed algorithm to obtain the Nash Equilibrium (NE) of the game was developed. Furthermore, based on the distributed algorithm, a network protocol for power control was presented. Dynamic spectrum sharing with multiple
strategic primary users (PUs) and SUs was investigated by using the noncooperative game in \cite{TVT11:P. Lin J. Jia Q. Zhang and M. Hamdi}, two cases under complete
and incomplete information assumptions were discussed.
The dynamic power control problem with interference constraints in CR networks was studied in \cite{TSP:M. Hong A. Garcia}. By enforcing the interference constraint through pricing, a non-cooperative game model was developed. A kind of
Generalized Nash Equilibrium (GNE) with the shared constraints, named as the interference equilibrium, was investigated. \par
The Stackelberg game, which is also referred to as the leader-follower
game, is a game in which the leader moves first and then the followers move sequentially. The problem is then transformed to find an optimal strategy for the leader, assuming that the followers
react in such a rational way that they optimize their
objective functions given the leader's actions \cite{LeadershipQuarterly:M. Hong A. Garcia}. In \cite{GameNets09:M. Bennis M. Le Treust S. Lasaulce M. Debbah and J. Lilleberg,TWC09:Y. Su and M. van der Schaar,TSP09:Y. Su and M. van der Schaar}\footnote{In \cite{TSP09:Y. Su and M. van der Schaar}, the Stackelberg equilibrium is a special case of the conjectural equilibrium.},
the Stackelberg game was applied for the multi-user
power control problem in interference channels. The Stackelberg game was used for power control in a decentralized multiple access
channel in \cite{TWC09:}. Moreover, in \cite{INFOCOM11:G. He S. Lasaulce and Y. Hayel}, it has been proved that compared to the standard non-cooperative power control game, the utilization of the Stackelberg game achieves performance improvement for both the individual and the global system. Distributed relay selection and power control
for multiuser cooperative communication
networks were addressed in \cite{TMC09:B. Wang Z. Han and K. J. R. Liu}.
In \cite{GLOBECOM10:S. Guruacharya D. Niyato E. Hossain and D.I. Kim}, the Stackelberg game was utilized to study the hierarchical competition in cellular networks that is comprised of the macrocells underlaid with femtocells.\par
As the Stackelberg game is defined for the cases in which a hierarchy of actions exists between players, it is a natural fit for the CR scenario. The Stackelberg game was employed to CR networks in \cite{Gamecomm07:M. Bloem T. Alpcan T. BaMar,CDC08:A. A. Daoud T. Alpcan S. Agarwal M. Alanyali}. A Stackelberg game model was proposed for frequency bands
in which a licensed user has priority over opportunistic cognitive
radios. In \cite{MobiHoc09:J. Zhang and Q. Zhang}, the Stackelberg game was applied for the utility-based cooperative CR networks.
In \cite{GLOBECOM09:Y. Li X. Wang M. Guizani}, the resource allocation in CR networks was studied by using the Stackelberg game to characterize the asymmetry of PUs and SUs. Allocation of under-utilized spectrum resources from PUs to multiple SUs was modeled
as the seller-buyer game.
Similar work can also be found in \cite{JSAC08:D. Niyato E. Hossain} though the authors did not claim the use of Stackelberg game explicitly. A decentralized
Stackelberg game formulation for power allocation was developed in \cite{ICASSP10: M. Razaviyayn Y. Morin and Z.-Q. Luo}.
Distributed optimization for CR
networks using the Stackelberg game was considered in \cite{ICCS10: Y. Xiao G. Bi and D. Niyato}.
Distributed power control method for SUs and optimal pricing for PU were obtained, and the algorithm for finding the optimal price was proposed. In \cite{ICC10: C. Yang J. Li}, the authors
focused on how the SU chooses its power level to obtain maximal cognitive network capacity with guaranteeing the
performance of the PU.
Power allocation in the down-link of the secondary system was considered by using the Stackelberg game in \cite{CAMAD11: N. Omidvar B. H. Khalaj}. Constraints such as protecting PUs and maximum power limitations of base stations (BSs) were considered.
Distributed power control for spectrum-sharing femtocell networks was investigated by using the Stackelberg game in \cite{ICC11: X. Kang Y.-C. Liang and H. K. Garg}.
The Stackelberg equilibrium (SE) was studied, and an effective distributed interference price
bargaining algorithm with guaranteed convergency was presented to achieve the equilibrium. \par
Recently, orthogonal frequency division multiplexing (OFDM) has been recognized as an attractive modulation candidate for CR systems. In practice, the efficient algorithm of allocating power to sub-carriers in OFDM-based PU network is also important.
However, most above mentioned works focus on the power control of the SU network, the hierarchic power allocation for OFDM-based PU network and SU network by using the Stackelberg game has not been extensively studied yet.
When the power control for the PU network and SU network are jointly considered, we should consider not only
the interference among SUs, but also the interference among PUs as well as the mutual interference between the PU network and the SU network. To meet quality of service (QoS) requirement of the PU precisely, the interference-to-signal ratio (ISR), which is defined as the ratio between the accumulated interference and the received signal power, should be less than a constant at the PU.
Then the power allocation of both PU network and SU network are tightly coupled.
In addition, the transmission from the primary transmitter to its receiver needs to be analyzed. The utility function of the PU takes the transmission merit, such as rate, into consideration.
 Due to the above reasons,
 the hierarchic power allocation algorithm is challenging especially when the PU network cannot acquire private information of the SU network.
 Even when the private information is available, it is difficult to design the time-efficient algorithm because of complexity of the game.
\par
In this paper,
the main contributions
are summarized as follows:
\begin{itemize}[]
\item
 A Stackelberg game is formulated to describe the priority of the power allocation for the PU network.
We analyze the mutual effect between power allocation for the PU network and that of the SU network in two aspects: ISR constraint and mutual interference between the PUs and SUs.
The former impacts the feasible power allocation set, and the latter influences the utility.
\item
When there is only one PU, the Stackelberg game can be written as an optimization problem that contains a non-cooperative sub-game. The sub-game
can be viewed as the power game of the SU network given the PU's power.
We analyze existence for the NE of the sub-game, and give a sufficient condition of uniqueness.
Moreover, an iterative algorithm, which converges to the NE, is presented for the general channel condition, and the closed-form solutions for the NE are derived in the perfectly symmetric channel.

\item Based on the Stackelberg game analysis, the hierarchic power allocation algorithms for the PU network and SU network are proposed. Considering availability of the private information for the SUs at the PU, two scenarios are investigated. When the private information is available and the perfectly symmetric channel conditions can be satisfied, the PU can allocate power by solving a specific optimization problem and the SU can allocate power analytically. Otherwise, the iterative distributed power allocation algorithms are presented. We also investigate convergence and effectiveness of the proposed iterative algorithms.
    \item  The extension to the multi-PU and multi-SU scenario is discussed, and we present an iterative distributed algorithm for the hierarchic power allocation.
\end{itemize}
\par
The reminder of the paper is structured as follows. In Section II, we introduce the system model under consideration,
and formulate the Stackelberg game. In Section III, the game analysis is performed.
In Section IV, the hierarchic power allocation methods for PU and SUs are proposed. Next, the numerical results are presented in Section V. We also discuss the extension to the multi-PU scenario in Section VI.
Finally, we conclude the paper in Section VII.
\section{System model and problem formulation}
\subsection{System model}
We consider a spectrum-sharing scenario
in which a PU system coexists with a SU system.
The PU system consists of a transceiver pair (i.e., PU) using OFDM. The SU system is an OFDM-based ad hoc network in which CR transceiver pairs (i.e., SUs) can simultaneously transmit with the PU. The PU is denoted as user 1 and the SUs are denoted as user 2,$\cdots$, user $L$, respectively, i.e., the PU set $\mathbb{P}=\{1\}$ and the SU set $\mathbb{S}=\{2,\cdots,L\}$.
It is assumed that the total number of OFDM sub-channels is $N$, and each sub-channel experiences flat fading.
The sampled signal on the $f$-th sub-channel at user $j$ is
$
y_{j}^{f}=\sqrt{P_j^fh_{j,j}^f}x_j^f+\sum\nolimits_{i \ne j \in \mathbb{P} \cup \mathbb{S}} \sqrt{P_i^fh_{i,j}^f}x_i^f+w_{i}^f,
$
where $P_j^f$ and $h_{i,j}^{f}$ denote the transmitted power of user $j$ and the channel coefficient between transmitter of user $i$ and receiver of user $j$ on the $f$-th sub-channel, respectively. $x_j^f$ is the transmitted symbol of user $j$ at sub-channel $f$
and is assumed to have unit energy.
$w_{i}^f$ is the additive white Gaussian noise (AWGN) with $w_{i}^f \sim \mathcal{CN}(0,N_{i}^{f})$. 
Each user has a limited power budget, i.e., $\sum\nolimits_{f=1}^{N}P_j^f \le P_j^{\max}$, $ \forall j \in \mathbb{P} \cup \mathbb{S}$.
Treating the interference as noise and assuming Gaussian signalling, the maximum rate that user $j$ can obtain on the $f$-th sub-channel can be expressed as
$R_j^f=\log\left(1+\frac{P_{j}^f|h_{j,j}^f|^2}{\sum_{i \ne j \in \mathbb{P} \cup \mathbb{S}}P_i^f|h_{i,j}^f|^2+N_{j}^{f}}\right)$ (nats/s/Hz).

\subsection{ Stackelberg game formulation}
We formulate the PU as the leader and the SUs as followers.
The PU first selects its transmission power by maximization of its utility, in which it tries to anticipate the SUs' reactions to its action.
And then, based on the PU's power, the SUs compete with each other to maximize its own rate by adjusting transmit power. The ISR constraint, $\frac{\sum\nolimits_{i \in \Omega}P_i^{f}|h_{i,1}^f|^2}{P_{1}^f|h_{1,1}^f|^2}\le \rho$ with $\rho$ being the ISR threshold,
needs to be satisfied to guarantee primary service\footnote{
We only need to guarantee that the power allocation in the stable state, i.e., the Stackelberg equilibrium (its definition will be given in the following) or the convergent outcomes of the iterative algorithm, should satisfy the ISR constraint.
}.
\par
Given the PU's transmit power, the SUs' non-cooperative sub-game can be mathematically formulated as
$
=\left\{\Omega ,\left\{\mathcal{S}_i\right\}_{i \in \Omega },\left\{u_i\right\}_{i \in \Omega }\right\},
$
where $\Omega=\mathbb{S}$ is the set of active players.
The set of admissible power allocation strategies for user $i$ is given by
$
\mathcal{S}_i=\Big\{\mathbf{P}_i=(P_{i}^1,P_{i}^2,\cdots,P_{i}^N):\sum\nolimits_{f = 1}^N P_i^f \le P_i^{\max };\
 \forall f \in \{1,2,\cdots,N\}, P_i^f \geq 0  \Big\}.
 $
The utility function of user $i$ is defined as
$u_i\left(\mathbf{P}_i,\mathbf{P}_{-i}\right)=\sum\nolimits_{f=1}^NR_{i}^f,$\footnote{The utility function can be defined in other forms,
i.e., the proposed framework is general enough to allow different definitions of the utility function. concerning the obtained conclusions, some are independent on the utility function definition and others can be adapted easily for new definitions of the utility function.}
where $\mathbf{P}_{-i}:=\left\{\mathbf{P}_{k}\right\}_{k \in \Omega/\{i\}}$.
\par
For the PU, if it can anticipate SUs' reactions to its action, we have the following problem
\begin{equation}\label{PU problem}
\mathop {\max }\limits_{\bf{P}_1} ~u_1=\sum\nolimits_{f=1}^{N}\log\left(1+\frac{P_{1}^f|h_{1,1}^f|^2}{\sum\nolimits_{i \in \Omega}P_i^{f*}|h_{i,1}^f|^2+N_{1}^{f}}\right)
\end{equation}
\[
\mbox{s.t.}
\sum\nolimits_{f = 1}^N {P_1^f }\le P_1^{\max } ,
P_1^f \ge 0,\\
\frac{\sum\nolimits_{i \in \Omega}P_i^{f*}|h_{i,1}^f|^2}{P_{1}^f|h_{1,1}^f|^2} \le \rho,
\]
where $\mathbf{P}_1=(P_1^1,P_1^2,\cdots,P_1^N)$, $\mathbf{P}_i^*=(P_i^{1*},P_i^{2*},\cdots,P_i^{N*})$ with $i \in \Omega$,
and $\left(\mathbf{P}_i^*,\mathbf{P}_{-i}^*\right)$ is the NE of $\mathcal{G}$ when $\mathbf{P}_1$ is given\footnote{The definition of NE will be given in the following Section.
(\ref{PU problem}) is the formulated Stackelberg game, where it contains the sub-game $\mathcal{G}$.
We should observe that
the ISR constraint is not considered in $\mathcal{G}$.
But as the ISR constraint is considered in (\ref{PU problem}), the solutions of the Stackelberg game comply with the ISR constraint.
}.
\section{Game analysis}
In this section, the existence, uniqueness, and solution for the NE of the sub-game $\mathcal{G}$ are analyzed. An iterative algorithm to obtain the NE of the sub-game is given. We also investigate the convergence of the iterative algorithm. Furthermore, the closed-form solutions for the NE are derived for the perfectly symmetric channel.
\par
\theoremstyle{definition} \newtheorem{Proposition}{Proposition}
\theoremstyle{definition} \newtheorem{lemma}{Lemma}
\theoremstyle{definition} \newtheorem{definition}{Definition}
First, for sub-game $\mathcal{G}$, its NE is defined as as follows:
\begin{definition}
$\left(\mathbf{P}_i^*,\mathbf{P}_{-i}^*\right)$ is the NE if
$u_i\left(\mathbf{P}_i^*,\mathbf{P}_{-i}^*\right) \ge u_i\left(\mathbf{P}_i,\mathbf{P}_{-i}^*\right)$ for all $\mathbf{P}_i \in \mathcal{S}_i$ and $i \in \Omega$.
\end{definition}
With respect to the existence of the NE for $\mathcal{G}$, we have the following proposition.
\begin{Proposition}\label{existence}
The sub-game $\mathcal{G}$ has at least one pure NE.
\end{Proposition}
\begin{IEEEproof}
Due to the page limitation, we give the sketch of proof.
First, $\forall \mathbf{P}, \mathbf{P}^{'} \in \mathcal{S}_i$, we have $\alpha\mathbf{P}+(1-\alpha)\mathbf{P}^{'}
\in \mathcal{S}_i$ ($\alpha \in [0,1]$), i.e., $\mathcal{S}_i$ is a convex set. Meanwhile, as $P_i^{\max}<\infty$, $\mathcal{S}_i \subseteq \mathbb{E}^{N}$ is closed and bounded, so it is compact. Next, $u_i\left(\mathbf{P}_i,\mathbf{P}_{-i}\right)$ is continuous in $\mathbf{P}_{-i}$. $\forall \tau \in \mathbb{R}$, we can prove that the upper contour set $\mathcal{U}_{\tau}=\left\{\mathbf{P}_i \in \mathcal{S}_i, u_i\left(\mathbf{P}_i,\mathbf{P}_{-i}\right)\ge \tau \right\}$ is convex. Consequently,
 $u_i\left(\mathbf{P}_i,\mathbf{P}_{-i}\right)$ is quasi-concave in $\mathbf{P}_i$.
 Using the Debreu-Fan-Glicksberg theorem \cite{book Game Theory: D. Fudenberg and J. Tirole}, the lemma can be proved.
\end{IEEEproof}
The uniqueness of the NE can be given by
\begin{Proposition}\label{uniqueness}
Define
\begin{eqnarray}
{\bf{M}}_{i,j} =
\left\{ \begin{array}{ll}
 - \mathop{ \max_{f \in [1,N]}} \left\{ \frac{|h_{ij}^f|^2}{|h_{ii}^f|^2}\frac{N_i^f +P_1^{f}|h_{1i}^f|^2+ \sum\nolimits_{l \in \Omega}|h_{li}^f|^2 P_l^{\max}}{N_j^f+P_1^{f}|h_{1j}^f|^2} \right\},
 & i \ne j;\\
1,& i = j.
\end{array} \right.
\end{eqnarray}
If ${\bf{M}}$ is a positive definite
matrix, $\mathcal{G}$ has a unique NE.
\end{Proposition}
\begin{IEEEproof}
Define $\Lambda_i\left(\mathbf{P}\right)=-\nabla_{\mathbf{P}_i}u_i\left(\mathbf{P}_i,\mathbf{P}_{-i}\right)$ with $\nabla_{\mathbf{P}_i}(\cdot)$ being the gradient vector with respect to $\mathbf{P}_i$,
and denote $\mathcal{S}=\mathcal{S}_2 \times \cdots \times \mathcal{S}_{|\Omega|+1}$ with a Cartesian structure. When ${\bf{M}}$ is a positive definite matrix, $\forall\mathbf{P}=(\mathbf{P}_2,\cdots,\mathbf{P}_{|\Omega|+1}), \mathbf{P'}= (\mathbf{P'}_2,\cdots,\mathbf{P'}_{|\Omega|+1})\in \mathcal{S}$, $\exists~ \alpha>0$ such that
$
\max_{i\in \Omega}\left\{\left(\mathbf{P}_i-\mathbf{P'}_i\right)\left[\Lambda_i\left(\mathbf{P}\right)-\Lambda_i\left(\mathbf{P'}\right)\right]\right\}\ge \alpha ||\mathbf{P}-\mathbf{P'}||_2^2,
$
where $||.||_2$ is the spectral norm. Consequently, $\mathcal{G}$ has a unique NE \cite{book Variational Inequalities: F. Facchinei and J.-S. Pang,WN02:T Alpcan T Basar R Srikant and E. Altman}.
\end{IEEEproof}
\emph{Remark: The conditions in Proposition \ref{uniqueness} can be viewed as the weak interference condition since $\frac{|h_{ij}^f|^2}{N_j^f+P_1^{f}|h_{1j}^f|^2}$ and $\frac{N_i^f +P_1^{f}|h_{1i}^f|^2+ \sum\nolimits_{l \in \Omega}|h_{li}^f|^2 P_l^{\max}}{|h_{ii}^f|^2}$ denote the interference level.}
\par
In the following, we give an iterative algorithm to obtain the NE.
The best response for user $i$ ($i \in \Omega$) can be expressed as
 \begin{eqnarray}\label{water-filling}
 P_i^f=BR_i\left(P_1^f,P_{-i}^{f}\right)
 =\left(\frac{1}{\mu_i}-\frac{P_1^f|h_{1,i}^f|^2+\sum\nolimits_{j \in \Omega/i}^{}P_j^f|h_{j,i}^f|^2+N_{i}^{f}}{|h_{ii}|^2}\right)^+,
 \end{eqnarray}
 where $P_{-i}^{f}(k)=\left\{P_j^f(k)\right\}_{j \in \Omega/i}$, $(\cdot)^+ =\max(\cdot,0)$, $\mu_i$ is a constant satisfying $\sum\nolimits_{f = 1}^N P_i^f \le P_i^{\max}$.
 Based on (\ref{water-filling}),
 an iterative distributed algorithm (Algorithm 1), which can converge to the NE, can be given.

In the algorithm, SU $i$ only has to obtain its own channel state, $h_{ii}$, and measure the aggregated interference it received,
$P_1^f|h_{1,i}^f|^2+\sum\nolimits_{j \in \Omega/i}^{}P_j^f(k)|h_{j,i}^f|^2$, therefore it can be implemented distributively.
\par
 Following the existing literature (such as \cite{TSP:M. Hong A. Garcia,JSAC07:K. W. Shum K.-K. Leung and C.. Sung}), sufficient conditions for the convergence of Algorithm 1 can be given by the following proposition.
 \begin{Proposition}
Define $c_{i,j}^f=|h_{i,j}^f|^2/|h_{j,j}^f|^2$, $[\mathbf{C}^f]_{i-1,j-1}=c_{i,j}^f$, $i \ne j \in \Omega$, and $[\mathbf{C}^f]_{i,i}=0$.
If $\forall f \in [1,\cdots,N]$, $||\mathbf{C}^f||<1$, where $||.||$ is any induced matrix norm with its corresponding vector norm being monotone, Algorithm 1 converges.
 \end{Proposition}
 \begin{IEEEproof}
 Please refer to \cite{TSP:M. Hong A. Garcia,JSAC07:K. W. Shum K.-K. Leung and C.. Sung}. The proof is omitted due to the page limitation.
 \end{IEEEproof}
Under a special circumstance, i.e., perfectly symmetric channel, we derive the closed-form solutions of NE.
\begin{Proposition}\label{Analytical NE}
When $|h_{i,j}^f|/|h_{j,j}^f|=|h_{j,i}^{f^{'}}|/|h_{i,i}^{f^{'}}|<1$, $N _i^f /|h_{ii}^f|^2=N _j^f /|h_{jj}^f|^2$ and $|h_{1i}^f{|}/|h_{ii}^f|=|h_{1j}^f{|}/|h_{jj}^f|$ for $f, f^{'} =1,\cdots,N $ and $i \ne j \in \Omega$, the perfectly symmetric channel conditions hold. Then, for $L=3$, the
NE of $\mathcal{G}$ has the following closed-form solutions\footnote{Without loss of generality, we assume $P_2^{\max } > P_3^{\max }$. $\sigma_f$ is only distinguished by the number of sub-channels in perfectly symmetric channel, the sub-channels can be re-numbered according to the strength of received PU'interference plus noise. Thus, it is also assumed that ${\sigma _1}\le{\sigma _2}\le\cdots\le{\sigma _N}$. Sub-carriers should be re-numbered at the beginning, and we need to recover the number of sub-carriers in the final.}.

\begin{eqnarray}\label{ne 2}
P_2^{f*} = \left\{
\begin{array}{ll}
t_1^* - \frac{ct_2^* + \sigma _f}{1 + c}, &  f \in [1,{k_2}];\\
t_1^* - \sigma _f, &  f \in [{k_2} + 1,{k_1}];\\
 0        ,&  f \in [{k_1} + 1,N],
\end{array} \right.
\end{eqnarray}

\begin{eqnarray}\label{ne 3}
P_3^{f*} = \left\{
\begin{array}{ll}
\frac{t_2^* - \sigma _f}{1 + c}, &  f \in [1,{k_2}];\\
 0 ,       &  f \in [{k_2} + 1,N],
\end{array} \right.
\end{eqnarray}
%
where
$
c=|h_{j,i}^f|^2|h_{i,i}^f|^{-2},$
$
{\sigma _f}=\left(N _i^f + P_1^f|h_{1i}^f{|^2} \right)|h_{ii}^f|^{-2},$
$
t_2^* = k_2^{-1}\left[(1 + c)P_3^{\max} + \sum\nolimits_{i = 1}^{k_2} {\sigma_i}\right],\label{1}
$
where $k_2$ can be found from
$
\varphi _{{k_2}}^2 < P_3^{\max } \le \varphi _{{k_2} + 1}^2
$
with
\begin{eqnarray}\label{3}
\varphi _k^2 =
\left\{ \begin{array}{ll}
 \frac{1}{{1 + c}}\sum\nolimits_{i = 1}^k {({\sigma _k} - {\sigma _i})} ,&1\le k \le N;\\
 \infty, &k = N+1.
\end{array} \right.
\end{eqnarray}
$
t_1^* = \frac{{P_2^{\max } + \sum\nolimits_{i = {k_2} + 1}^{{k_1}} {{\sigma _i}}  + \frac{1}{{1 + c}}\sum\nolimits_{i = 1}^{{k_2}} {(ct_2^* + {\sigma _i})} }}{{{k_1}}},
$
where $k_1 = k_2$ when $P_2^{\max} \le \varphi_{k_2 + 1}^1$; Otherwise, $k_1$ is the solution of
$
\varphi _{{k_1}}^1 < P_2^{\max } \le \varphi _{{k_1} + 1}^1
$
and $\varphi _k^1$ is defined as
\begin{eqnarray}
\varphi _k^1 =
\left\{ \begin{array}{l}
 \sum\nolimits_{i = k_2 + 1}^k ({\sigma _k} - {\sigma _i}) +\frac{1}{1 + c}\sum\nolimits_{i = 1}^{{k_2}}\Big( (1 + c){\sigma _k} \\
  - {\sigma _i} - ct_2^* \Big), k \in [{k_2} + 1,N];\\
 \infty,~~~~~~~~~~~k=N+1.
\end{array} \right.
\end{eqnarray}
\end{Proposition}
\begin{IEEEproof}
Let $|\Omega|$ be the cardinality of the set $\Omega$.
Since $u_i\left(\mathbf{P}_i,\mathbf{P}_{-i}\right)$ is concave on $\mathbf{P}_i$, using the KKT conditions \cite{book Convex Optimization: S. Boyd and L. Vandenberghe}, $\left(\mathbf{P}_2,\cdots,\mathbf{P}_{|\Omega|+1}\right)$ is the NE if and only if there are non-negative $\{\mu_i\}$ satisfying
 \begin{eqnarray}
\lefteqn{
\frac{\partial u_i\left(\mathbf{P}_i,\mathbf{P}_{-i}\right)}{\partial P_{i}^{f}}
=\left[P_{i}^{f}+\frac{N _i^f + P_1^f|h_{1i}^f{|^2} }{|h_{ii}^f|^2}+\frac{\sum\nolimits_{j \ne i \in \Omega}P_{j}^{f}|h_{ji}^f|^2}{|h_{ii}^f|^2}\right]^{-1}}
\label{kkt results}\\
&=&\left[P_{i}^{f}+\sigma^f+c\sum\nolimits_{j \ne i \in \Omega}P_{j}^{f}\right]^{-1}
\left\{ \begin{array}{l}
=\mu_i, ~ P_{i}^{f}>0;\\
\le \mu_i,~P_{i}^{f}=0.
\end{array} \right.
\end{eqnarray}
Consequently,
let
$
\tau_{r}^{k}=\frac{1}{1-c}\left(\frac{1+(|\Omega|-1-r+k)c}{\lambda_k}-c\sum\nolimits_{j=1}^{|\Omega|-r+k}\frac{1}{\lambda_j}\right)
$
with $\lambda_1 \le \cdots \le \lambda_{|\Omega|}$, each NE is of the form as
\begin{eqnarray}
P_{k+1}^{f}=
\left\{ \begin{array}{l}
\frac{1}{1+(|\Omega|-1)c}(\tau_{k}^{k}-\sigma _f), \sigma _f<\tau_{|\Omega|}^{|\Omega|};\\
\frac{1}{1+(|\Omega|-1-r+k)c}(\tau_{r}^{k}-\sigma _f), \tau_{|\Omega|}^{|\Omega|+k+1-r}
 \le\sigma _f<\tau_{|\Omega|}^{|\Omega|+k-r}, r \in [k+1,|\Omega|];\\
0, \tau_{|\Omega|}^{k} \le \sigma _f.
\end{array} \right.
\end{eqnarray}
For user $(|\Omega|+1)$, we have
\begin{eqnarray}
\lefteqn{
\sum\nolimits_{f=1}^{N}P_{|\Omega|+1}^{f}=\sum\nolimits_{\sigma _f<\tau_{|\Omega|}^{|\Omega|}}P_{|\Omega|+1}^{f}
}
\nonumber \\
&=&\frac{1}{1+(|\Omega|-1)c}\sum\nolimits_{\sigma _f<\tau_{|\Omega|}^{|\Omega|}}\left(\tau_{|\Omega|}^{|\Omega|}-\sigma _f\right)
\le P_{|\Omega|+1}^{\max}.
\end{eqnarray}
When the equality holds, we have $\tau_{|\Omega|}^{|\Omega|*}=\frac{(1+(|\Omega|-1)c)P_{|\Omega|+1}^{\max}+\sum\nolimits_{f=1}^{k_{|\Omega|}}\sigma_f}{k_{|\Omega|}},$ where $k_{|\Omega|}$ is given by
$\phi_{k_{|\Omega|}}^{|\Omega|}<P_{|\Omega|+1}^{\max} \le \phi_{k_{|\Omega|}+1}^{|\Omega|}$ and $\phi_{k}^{|\Omega|}=\frac{1}{1+(|\Omega|-1)c}\sum\nolimits_{f=1}^{k}(\sigma_k-\sigma_f)$. Consequently, the equilibrium power allocation for user $(|\Omega|+1)$ is given by
\begin{eqnarray}
P_{|\Omega|+1}^{f*} = \left\{ \begin{array}{ll}
\frac{\tau_{|\Omega|}^{|\Omega|*}-\sigma_f}{1+(|\Omega|-1)c},  &f \in [1,k_{|\Omega|}];\\
0,  &f \in [k_{|\Omega|}+1,N].
\end{array} \right.
\end{eqnarray}
$\tau_{|\Omega|-1}^{|\Omega|-1}=\frac{1+(|\Omega|-1)c}{1+(|\Omega|-2)c}\tau_{|\Omega|}^{|\Omega|-1}-\frac{c}{1+(|\Omega|-2)c}\tau_{|\Omega|}^{|\Omega|}$, then regarding user $|\Omega|$,
\begin{eqnarray}
\lefteqn{
\sum\nolimits_{f=1}^{N}P_{|\Omega|}^{f}=\frac{1}{1+(|\Omega|-1)c}\sum\nolimits_{\sigma _f<\tau_{|\Omega|}^{|\Omega|}}\left(\tau_{|\Omega|-1}^{|\Omega|-1}-\sigma _f\right)
}\nonumber \\
&+&\frac{1}{1+(|\Omega|-2)c}\sum\nolimits_{\tau_{|\Omega|}^{|\Omega|}\le \sigma _f<\tau_{|\Omega|}^{|\Omega|-1}}\left(\tau_{|\Omega|}^{|\Omega|-1}-\sigma _f\right)
\le P_{|\Omega|}^{\max}.
\end{eqnarray}
Utilizing the equality, we get
$
\tau_{|\Omega|}^{(|\Omega|-1)*}=
\bigg(P_{|\Omega|}^{\max}+\frac{\sum\nolimits_{f=k_{|\Omega|}}^{k_{|\Omega|-1}}\sigma _f}{1+(|\Omega|-2)c}
+\frac{\sum\nolimits_{f=1}^{k_{|\Omega|}}
(\frac{c\tau_{|\Omega|}^{|\Omega|*}}{1+(|\Omega|-2)c}+\sigma _f)}{1+(|\Omega|-1)c}\bigg)
\left(\frac{1+(|\Omega|-2)c}{k_{|\Omega|-1}}\right),
$
where $k_{|\Omega|-1}$ is derived by
\begin{eqnarray}
\left\{ \begin{array}{ll}
k_{|\Omega|-1} = {k_{|\Omega|}}{,^{}} &P_{|\Omega|}^{\max} \le \phi_{k_{|\Omega|}+1}^{|\Omega|-1};\\
\phi_{k_{|\Omega|-1}}^{|\Omega|-1} < P_{|\Omega|}^{\max} \le\phi_{k_{|\Omega|-1}+1}^{|\Omega|-1}, &\mathrm{otherwise}.
\end{array} \right.
\end{eqnarray}
with
\begin{eqnarray}
\lefteqn{
\phi_{k}^{|\Omega|-1}= \sum\nolimits_{f=k_{|\Omega|}+1}^{k}\frac{\sigma_k-\sigma_f}{1+(|\Omega|-2)c}+\sum\nolimits_{f=1}^{k_{|\Omega|}}\frac{1}{1+(|\Omega|-2)c}
}
\nonumber\\
&\times&\left(\frac{1+(|\Omega|-1)c}{1+(|\Omega|-2)c}\sigma_k-\sigma_f+\frac{c}{1+(|\Omega|-2)c}\tau_{|\Omega|}^{|\Omega|*}\right).
\end{eqnarray}
Then
\begin{eqnarray}
P_{|\Omega|}^{f*} = \left\{ \begin{array}{ll}
\frac{\tau_{|\Omega|}^{(|\Omega|-1)*}}{1+(|\Omega|-2)c}-\frac{\frac{c\tau_{|\Omega|}^{|\Omega|*}}{1+(|\Omega|-2)c}+\sigma_f}{1+(|\Omega|-1)c}, &f \in [1,k_{|\Omega|}];\\
\frac{\tau_{|\Omega|}^{(|\Omega|-1)*}-\sigma_f}{1+(|\Omega|-2)c}, &f \in [k_{|\Omega|}+1,k_{|\Omega|-1}];\\
0, &f \in [k_{|\Omega|-1}+1,N].
\end{array} \right.
\end{eqnarray}
As $|\Omega|=2$, we arrive at the proposition, which completes the proof.
\end{IEEEproof}
\emph{Remark: The above proposition is for the 2-SU scenario, however, following the proof of this proposition, the closed-form solutions for the multi-SU scenario can be obtained similarly. Using Proposition \ref{Analytical NE}, the power for SUs in the perfectly symmetric channel can be allocated analytically with simple computation. Moreover, if we suppose that $\{P_i^{\max}\}_{i \in \Omega}$ is known at user $i$ ($i \in \Omega$), user $i$ ($i \in \Omega$) only needs to obtain $c$ (i.e. $h_{j,i}^f$ and $h_{i,i}^f$) and measure the received interference from PU, $P_{1}^f|h_{1i}|^2$. Thus Proposition \ref{Analytical NE} can be distributively applied. }

\par
 Equations (\ref{ne 2}) and (\ref{ne 3}) as well as Algorithm 1 can be used to obtain the NE of $\mathcal{G}$ in the 2-SU scenario. When the perfectly symmetric channel conditions hold, the analytical solutions are given in (\ref{ne 2}) and (\ref{ne 3}); Otherwise, Algorithm 1 can find the solution for the general case.
\section{Power allocation algorithm}
In this section, we consider the hierarchic power allocation for the PU and SUs. If the PU can acquire the additional information about the SUs to anticipate SUs' reactions to its action, we propose an analytical power allocation algorithm. Otherwise, the iterative power allocation algorithms are developed.
\subsection{Analytical power allocation algorithm}
The definition of the SE is given by
\begin{definition}\label{SEdefination}
$(\mathbf{P}_1^{*},\hat{\mathbf{P}}_i^{*},\hat{\mathbf{P}}_{-i}^{*})$ is a SE for the proposed
Stackelberg game when it satisfies
\begin{enumerate}[]
\item
$u_i\left(\mathbf{P}_1^{*},\hat{\mathbf{P}}_i^{*},\hat{\mathbf{P}}_{-i}^{*}\right)\ge u_i\left(\mathbf{P}_1^*,\mathbf{P}_i,
\hat{\mathbf{P}}_{-i}^{*}\right), \forall i \in \Omega, \mathbf{P}_i \in \mathcal{S}_i$.
\item
$u_1\left(\mathbf{P}_1^{*},\hat{\mathbf{P}}_i^{*},\hat{\mathbf{P}}_{-i}^{*}\right)\ge u_1\left(\mathbf{P}_1,\mathbf{P}_i^{*},
\mathbf{P}_{-i}^{*}\right)$ for any feasible $\mathbf{P}_1$.
\end{enumerate}
\end{definition}
\emph{Remark: In the definition, inequality 1) implies that $(\hat{\mathbf{P}}_i^{*},\hat{\mathbf{P}}_{-i}^{*})$ is the NE of $\mathcal{G}$ given $\mathbf{P}_1^{*}$. As $\left(\mathbf{P}_i^*,\mathbf{P}_{-i}^*\right)$ denotes the NE of $\mathcal{G}$ given $\mathbf{P}_1$, we have an equivalent definition: $(\mathbf{P}_1^{*},Ne(\mathbf{P}_1^{*}))$ is a SE if $u_1\left(\mathbf{P}_1^{*},Ne(\mathbf{P}_1^{*})\right)\ge u_1\left(\mathbf{P}_1,Ne(\mathbf{P}_1)\right)$ for any feasible $\mathbf{P}_1$, where $Ne(x)$ denotes the NE of $\mathcal{G}$ given $\mathbf{P}_1=x$.}
\par
The following lemma gives the solution of the SE for the proposed Stackelberg game.
\begin{lemma}\label{SE}
The SE of the proposed Stackelberg game can be obtained as follows: 1) Solving (\ref{PU problem}) to obtain $\mathbf{P}_1^{*}$. 2) Let $\mathbf{P}_1=\mathbf{P}_1^{*}$, solving the NE of $\mathcal{G}$, $(\hat{\mathbf{P}}_i^{*},\hat{\mathbf{P}}_{-i}^{*})$. Then, $(\mathbf{P}_1^{*},\hat{\mathbf{P}}_i^{*},\hat{\mathbf{P}}_{-i}^{*})$ is a SE.
\end{lemma}
\begin{IEEEproof}
$(\hat{\mathbf{P}}_i^{*},\hat{\mathbf{P}}_{-i}^{*})$ is the NE solutions of $\mathcal{G}$ given $\mathbf{P}_1^{*}$, so we have
 \begin{eqnarray}\label{SEproof1}
 u_i\left(\mathbf{P}_1^{*},\hat{\mathbf{P}}_i^{*},\hat{\mathbf{P}}_{-i}^{*}\right)\ge u_i\left(\mathbf{P}_1^*,\mathbf{P}_i,
\hat{\mathbf{P}}_{-i}^{*}\right), \forall i \in \Omega, \mathbf{P}_i \in \mathcal{S}_i.
\end{eqnarray}
 Furthermore,
since $\mathbf{P}_1^{*}$ is the optimal solution of (\ref{PU problem}), then
 \begin{eqnarray} \label{SEproof2}
 u_1\left(\mathbf{P}_1^{*},\hat{\mathbf{P}}_i^{*},\hat{\mathbf{P}}_{-i}^{*}\right)=u_1\left(\mathbf{P}_1^{*},Ne(\mathbf{P}_1^{*})\right)
 \ge
 u_1\left(\mathbf{P}_1^{},Ne(\mathbf{P}_1)\right)
 =u_1\left(\mathbf{P}_1^{},\mathbf{P}_i^{*},\mathbf{P}_{-i}^{*}\right)
  \end{eqnarray}
  for any feasible $\mathbf{P}_1$.
Combing (\ref{SEproof1}) and (\ref{SEproof2}), we claim that $(\mathbf{P}_1^{*},\hat{\mathbf{P}}_i^{*},\hat{\mathbf{P}}_{-i}^{*})$ is a SE, which competes the proof.
\end{IEEEproof}
\par
Based on Lemma \ref{SE}, we get the analytical power allocation method. First, the PU obtains the optimal power allocation, $\mathbf{P}_1^{*}$, by solving (\ref{PU problem}). Then, SUs allocate the power according to the NE of $\mathcal{G}$, $(\hat{\mathbf{P}}_i^{*},\hat{\mathbf{P}}_{-i}^{*})$, given $\mathbf{P}_1=\mathbf{P}_1^{*}$.
 For the 2-SU scenario with perfectly symmetric channels, substituting (\ref{ne 2}) and (\ref{ne 3}) into (\ref{PU problem}),
 the PU problem becomes a convetional non-convex optimization problem.
By solving the problem\footnote{The
PU should know $c$, $h_{1i}^f$, $h_{ii}^f$, $h_{i1}^f$, $P_i^{\max}$ ($i \in \Omega$, $f=1,\cdots,N$) and its own channel state $h_{11}^f$ to solve the problem numerically.}, we obtain the optimal power allocation strategy of PU, $\mathbf{P}_1^{*}$. Replacing $\mathbf{P}_1$ by $\mathbf{P}_1^{*}$ in (\ref{ne 2}) and (\ref{ne 3}), we get the NE of $\mathcal{G}$ given the optimal power allocation of PU, denoted by $(\hat{\mathbf{P}}_2^{*},\hat{\mathbf{P}}_3^{*})$. Then, SUs allocate the power according to $\hat{\mathbf{P}}_2^{*}$ and $\hat{\mathbf{P}}_3^{*}$, respectively. Observe that $(\mathbf{P}_1^{*},\hat{\mathbf{P}}_2^{*},\hat{\mathbf{P}}_3^{*})$ is
the SE of the Stackelberg game according to Lemma \ref{SE}.
\subsection{Iterative power allocation algorithm}
If the private information of the SUs is unknown to the PU, the PU cannot set an optimal power level by solving the non-convex optimization problem even under the perfectly channel conditions in Proposition \ref{Analytical NE}. Alternatively, the iterative algorithms are needed to identify the power level.
\par
The outcomes of the iterative algorithms are not the SE solution.
To play SE, the PU must have the ability to anticipate the SUs' reactions to its action. However, it is impossible to exactly anticipate the SUs' reactions to the PU's action when the PU cannot obtain the private information about the SUs. The PU should know the SUs' private information such as the strategy set (please refer to footnote 6 to find the exact information needed) to anticipate the SUs' reactions to its action. Although SE can be viewed as a special case of conjectural equilibrium (CE) \cite{TSP09:Y. Su and M. van der Schaar}, CE assumes that the foresighted user knows its stationary interference and the first derivatives with respect to the allocated power (ISR constraint is not considered in \cite{TSP09:Y. Su and M. van der Schaar}). Hence, no algorithms can derive the SE solution in the case that the PU cannot obtain the private information about the SUs, especially when the ISR constraint is considered.
\par
The PU sets an initial power level in Step 1. In each iteration, based on PU's power allocation in the former iteration, SUs allocate their power levels $\left\{\mathbf{P}_{i}(n)=\left(P_i^1(n),\cdots,P_i^N(n)\right)\right\}_{i \in \mathbb{S}}$ according to the NE of the SUs' sub-game by using Proposition \ref{Analytical NE} or Algorithm 1. Given the
novel power levels of the SUs, the PU updates its power by maximizing its utility under total power and interference constraints\footnote{Please refer to (\ref{PU problem}). To some extent, the ISR constraint is imposed on PU network in the iterative algorithm.
In \cite{ICC11: X. Kang Y.-C. Liang and H. K. Garg}, the interference constraint has been imposed on PU to decrease the complexity of the power allocation algorithms. Here we impose ISR constraint on PU network for the similar reason.}, i.e., $P_1^f(n+1)$ is the solution of the following convex optimization problem,
\begin{equation}\label{optforPU in IPASUPU}
\mathop {\max }\limits_{\bf{P}_1} ~u_1=\sum\nolimits_{f=1}^{N}\log\left(1+\frac{P_{1}^f|h_{1,1}^f|^2}{I^f(n)+N_{1}^{f}}\right)
\end{equation}
\[
\mbox{s.t.}
\sum\nolimits_{f = 1}^N {P_1^f} \le P_1^{\max } ,
P_1^f \ge 0,
\frac{I^f(n)}{P_{1}^f|h_{1,1}^f|^2} \le \rho,
\]
where
$
I^f(n)=\sum\nolimits_{i \in \mathbb{S}}P_i^{f}(n)|h_{i,1}^f|^2
$
is the received interference at the PU. The ISR constraint $\frac{I^f(n)}{P_{1}^f|h_{1,1}^f|^2}\le \rho$ in (\ref{optforPU in IPASUPU}) is equivalent to a minimal power constraint $\frac{I^f(n)}{\rho|h_{1,1}^f|^2}\le P_{1}^f$. Consequently,
it can be solved by a 2-step algorithm. The minimal power to meet the ISR constraint is first allocated to each sub-channel, i.e.,  we allocate $\frac{I^f(n)}{\rho|h_{1,1}^f|^2}$ for sub-channel $f$; Then, subtracting the allocated power from $P_{1}^{\max}$ and allocating the remaining power to the sub-channels by using water-filling method. The iteration continues until convergence. We observe
that the PU only needs to know its own channel information, $h_{11}^f$, and the received interference,
$I^f(n)$. The specific distributed power allocation algorithm is described in Algorithm 2.


\emph{Remark: When the private information of the SUs (followers) cannot be acquired by the PU (leader), the PU has no information at the beginning and it cannot anticipate the interference from the SUs with respect to its own power allocation, the only thing it can do is to randomly set an initial feasible power allocation. Then according to the PU's power allocation, the SUs play their sub-game to obtain the power allocations.
Next, define the $n$-th ($n=1,2...$) round as \lq\lq the PU allocates its power $\mathbf{P}_{1}(n)$, and the SUs allocate the power $\{\mathbf{P}_{i}(n)|i \in \Omega\}$ subsequently\rq\rq. In the $n$-th round, the PU can only know the interference of the SUs with respect to the PU's former power allocation (power allocation in the former round), i.e., $I^f(n-1)$ (history information of the interference and can be obtained by measuring the total interference it received), it cannot exactly anticipate the interference of the SUs with respect to the PU's allocation in the same round, i.e., $I^f(n)$ (future information of the interference), so it can only allocate the power by utilizing the history information $I^f(n-1)$. Then, based on the PU's power allocation, the SUs play their sub-game to obtain the power allocations in the same round. In addition, the ISR constraint should be considered in the power allocation.}
\par
\emph{
Due to the condition that the PU cannot obtain the private information
about the SUs, the PU cannot exactly
anticipate the future information of the interference\footnote{Based on the history information of the interference, the PU may predict the future information of the interference by using prediction methods, but it is not exact prediction.} and it can only utilize the
history information of the interference. In conclusion, the unavailability of the private information
and the ISR constraint lead to Algorithm 2. There are many methods to utilize the history information, we choose the simplest one in our algorithm.
}
\par
Regarding the convergence of Algorithm 2, we have the following lemma.
\begin{lemma}\label{convergence of A2}
When $P_i^{\max}$, $h_{ij}$, and $N_i$ ($i , j \in \mathbb{S} \cup \mathbb{P}$) are fixed, there exists a constant $\xi>0$, and when $\eta <\xi$,
Algorithm 2 converges.
\end{lemma}
 \begin{IEEEproof}
 Denote
 $$\chi(\mathbf{P}_{1}(n))=\left[\frac{I^f(n)}{\rho|h_{1,1}^f|^2}+\left(\lambda-\frac{I^f(n)+N_1^f}{|h_{11}|^2}\right)^+\right]_{f=1}^{N},
 $$ where $\Big[x_i\Big]_{i=1}^{n}=(x_1,\cdots,x_n)$. Then
 \begin{eqnarray} \label{converge1}
 \mathbf{P}_{1}(n+1)
 =(1-\eta)\mathbf{P}_{1}(n)+\eta\chi(\mathbf{P}_{1}(n))
 :=F\left(\mathbf{P}_{1}(n)\right).
 \end{eqnarray}
 First, $\forall\mathbf{P}_{1}^{(1)} \ne ~\mathbf{P}_{1}^{(2)}$ in PU's feasible power set, as $\sum\nolimits_{f = 1}^N P_i^f \le P_i^{\max }$ for $i \in \mathbb{P} \cup \mathbb{S}$, $\exists ~\beta>0$ satisfies
 \begin{eqnarray} \label{converge2}
\left(\mathbf{P}_{1}^{(1)}-\mathbf{P}_{1}^{(2)}\right)\left[\chi(\mathbf{P}_{1}^{(1)})-\chi(\mathbf{P}_{1}^{(2)})\right]^T
  \ge-\beta ||\mathbf{P}_{1}^{(1)}-\mathbf{P}_{1}^{(2)}||_2^2.
 \end{eqnarray}
 Next, from (\ref{converge1}), we get
 \begin{eqnarray}
 \lefteqn{
 \left(\mathbf{P}_{1}^{(1)}-\mathbf{P}_{1}^{(2)}\right)\left[F\left(\mathbf{P}_{1}^{(1)}\right)-F\left(\mathbf{P}_{1}^{(2)}\right)\right]^{T}
 =(1-\eta)\left(\mathbf{P}_{1}^{(1)}-\mathbf{P}_{1}^{(2)}\right)\left(\mathbf{P}_{1}^{(1)}-\mathbf{P}_{1}^{(2)}\right)^{T}
 }
  \nonumber\\
 &+&\eta\left(\mathbf{P}_{1}^{(1)}-\mathbf{P}_{1}^{(2)}\right)\left[\chi(\mathbf{P}_{1}^{(1)})-\chi(\mathbf{P}_{1}^{(2)})\right]^{T}
 \stackrel{(a)}{\ge} \left[1-(1+\beta)\eta\right]||\mathbf{P}_{1}^{(1)}-\mathbf{P}_{1}^{(2)}||_2^2,
 \end{eqnarray}
 where $(a)$ holds since (\ref{converge2}).
 On the other hand, $\exists ~\theta>0$,
 $
 ||\chi(\mathbf{P}_{1}^{(1)})-\chi(\mathbf{P}_{1}^{(2)})||_2 \le \theta ||\mathbf{P}_{1}^{(1)}-\mathbf{P}_{1}^{(2)}||_2.
 $
 Consequently, we derive
\begin{eqnarray}
 \lefteqn{
 \left(\mathbf{P}_{1}^{(1)}-\mathbf{P}_{1}^{(2)}\right)\left[F\left(\mathbf{P}_{1}^{(1)}\right)-F\left(\mathbf{P}_{1}^{(2)}\right)\right]^{T}
 }\nonumber\\
 &\ge& (1-(1+\beta)\eta)\theta^{-2}||\chi(\mathbf{P}_{1}^{(1)})-\chi(\mathbf{P}_{1}^{(2)})||_2^2.
  \end{eqnarray}
  When $\eta<{(1+\beta)^{-1}}$, $F(\cdot)$ is co-coercive with constant $[1-(1+\beta)\eta]\theta^{-2}$.
 Then,
 applying Th. 12.1.8 in \cite{book Variational Inequalities: F. Facchinei and J.-S. Pang}, if $\eta <2[1-(1+\beta)\eta]\theta^{-2}$, i.e., $\eta <2[2(1+\beta)-\theta^2]^{-1}$, the iterative algorithm converges. In conclusion if $\eta<\min\left\{(1+\beta)^{-1},2[2(1+\beta)-\theta^2]^{-1}\right\}=(1+\beta)^{-1}:=\xi$,
 the iterative algorithm converges, which completes the proof.
\end{IEEEproof}
\emph{Remark: The upper bound of convergent step-size for Algorithm 2 is fixed.
If the algorithm does not converge with a certain step-size, we can choose smaller step-size to make the algorithm converge. Lemma \ref{convergence of A2} guarantees the existence of such convergent step-size.}
\par
In \cite{TSP09:Y. Su and M. van der Schaar}, conjecture-based rate maximization (CRM) algorithms are developed even if the foresighted user has no a priori knowledge of its competitors' private information. The CRM algorithm can achieve better performance than NE\footnote{Observe that the CRM algorithm cannot derive the SE.}. However, there are shortcomings of CRM algorithm: 1) It is not guaranteed to converge to a CE. 2) It cannot be utilized for the scenarios in which multiple foresighted users coexist. 3) The number of frequency bins should be sufficiently large. In contrast, there are no constraints on the number of frequency bins in our proposed algorithm, and it has guaranteed convergence performance. Moreover, our proposed algorithm can be extended to the multi-leader case (see Section VI). Finally, no ISR constraints are considered in \cite{TSP09:Y. Su and M. van der Schaar}\footnote{The system model considered in \cite{TSP09:Y. Su and M. van der Schaar} is the interference channel.}. As explained in the paper, the ISR constraint will greatly couple the power allocations of the PU (leader) with the power allocations of the SUs (followers). That is to say, the CRM algorithm cannot be applied under our system model, where the ISR constraint should be considered. In a word, we deal with a more complicated problem in this paper.
\par
In Algorithm 2, the PU waits for the convergence of
the power profiles of the SUs (Step 2), it then updates its power. It will be time-consuming especially when the number of SUs is large. For the purpose of further improving time efficiency, we propose the asynchronous algorithm in Algorithm 3.

\par
\emph{Remark: The PU asynchronously updates its power allocation in Algorithm 3. It does not need to wait for the convergence of SUs' power allocation. Consequently, it is more time-efficient.}
\par

\section{Numerical results}
In this section, we perform simulations to verify our analysis. The convergence of the iterative algorithm as well as the rate performance for analytical and iterative algorithms are given numerically in this section. In the simulation,
the channel coefficients are modeled as
independently circular symmetric Gaussian distributed random variables for the convenience of illustration.
We also assume that the channels do not change during one implementation of the algorithm and \lq\lq average\rq\rq (e.g., average power, average rate)
is taken over $10^4$ channel realizations.
\par
\begin{figure}[]
\centering
\includegraphics[width=4.5in]{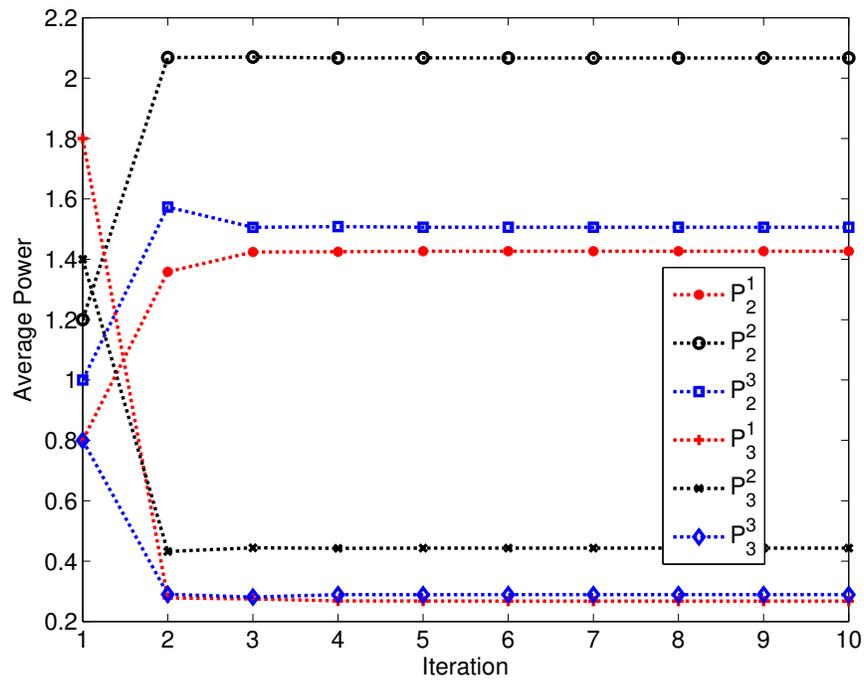}
\caption{Power allocation of the SUs by using Algorithm 1 in the perfectly symmetric channel case, there are 3 sub-carriers and PU's power is $P_1=[7~ 1~3 ].$}
\label{sim fig1}
\end{figure}

First, we compare the analytical solutions (\ref{ne 2}) and (\ref{ne 3}) with Algorithm 1 in the perfectly symmetric channel case.
In the simulations, we set $N=3$, $P_1=[7 ~1~ 3]$, $P_2^{\max}=5$, $P_3^{\max}=1$, $N_2=N_3=[0.5~ 0.5 ~0.5]$, $h_{22}=h_{33}\sim \mathcal{CN}\left(0, [1~ 1~ 1]\right)$,
$h_{12}=h_{13}\sim \mathcal{CN}\left(0,[\sqrt{0.2}~ \sqrt{0.3}~ \sqrt{0.4}]\right)$, and
$h_{23}=h_{32}=0.5\times h_{22}$ (i.e., $c=0.25$). Using (\ref{ne 2}) and (\ref{ne 3})\footnote{Sub-carriers should be re-numbered before using (\ref{ne 2}) and (\ref{ne 3}), and we need to recover the number of sub-carriers in the final.}, we obtain the average NE power $\mathbf{P}_2^{*} =[  1.4242  ~  2.0709 ~   1.5049]$, $\mathbf{P}_3^{*} = [ 0.2676  ~  0.4432  ~  0.2892]$.
Fig. \ref{sim fig1} shows the results of Algorithm 1. Observe that Algorithm 1 converges to the same results as analytical solutions since the $5$-th iteration.
\par
Next, we evaluate the convergence performance of the iterative hierarchic power allocation algorithms.
The inner iteration for Algorithm 2 (iteration for Algorithm 1) is set to be $10$. Iteration denotes the number of the outer iterations in Algorithm 2. In Algorithm 3, we let $\tau_k=3\times k$.
\par
\begin{figure}[]
\centering
\includegraphics[width=5in]{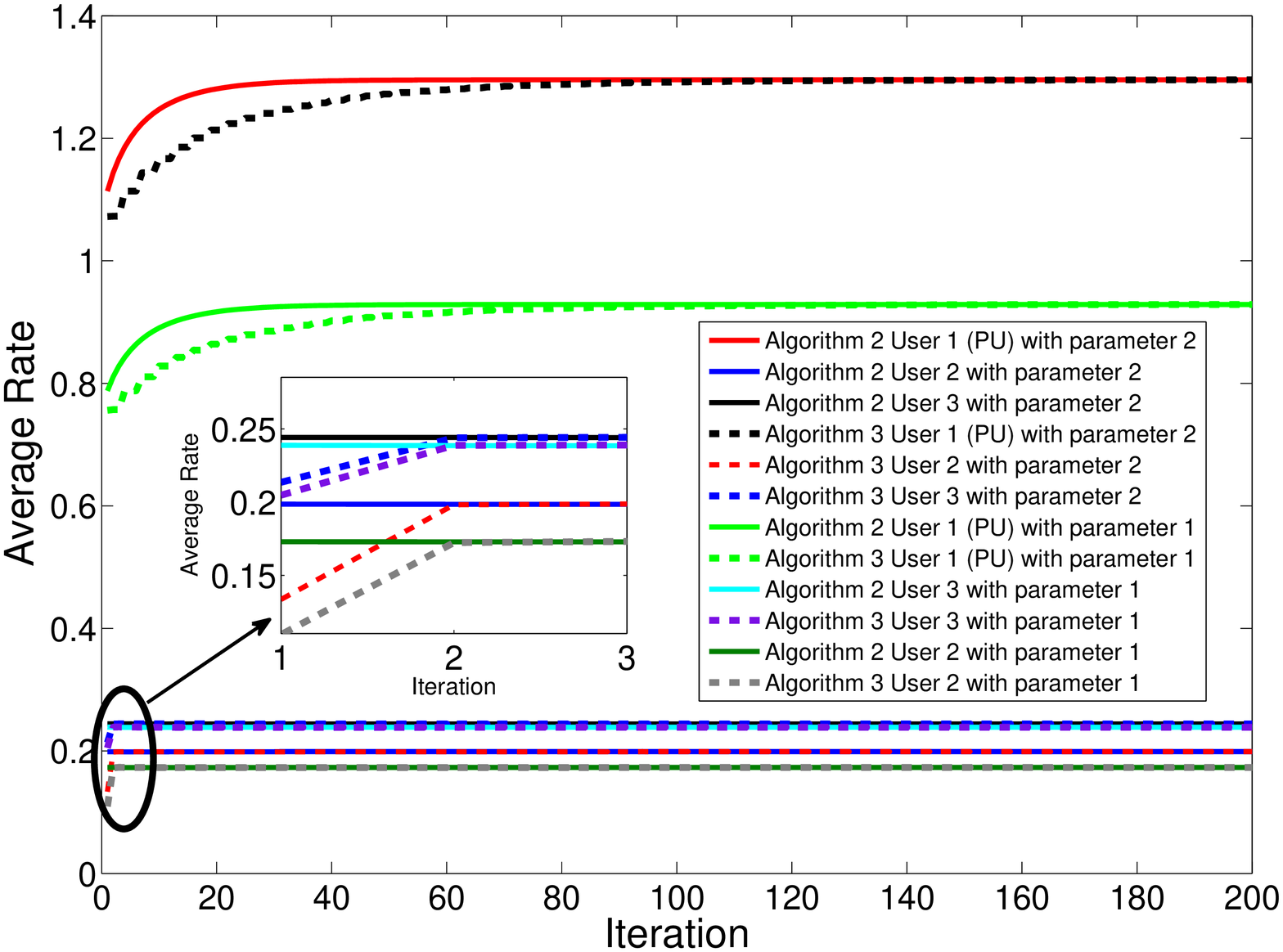}
\caption{Convergence performance of Algorithm 2 and Algorithm 3 with convergence performance of Algorithm 2 and Algorithm 3 with different channel parameters. Channel parameters 1: $\tilde{h}_{12}\sim \mathcal{CN}\left(0,[0.7~ 0.5~ 0.6]\right)$,
$\tilde{h}_{13}\sim \mathcal{CN}(0,[0.5~ 0.5~ 0.7])$,
$\tilde{h}_{21}\sim\mathcal{CN}(0,[0.4~ 0.5~ 0.6])$,
$\tilde{h}_{31}\sim\mathcal{CN}(0,[0.5~ 0.5~ 0.4])$,
$\tilde{h}_{23}\sim\mathcal{CN}(0,[0.5~ 0.5~ 0.5])$,
$\tilde{h}_{32}\sim\mathcal{CN}(0,[0.5~ 0.5~ 0.5])$,
$\tilde{h}_{11}\sim\mathcal{CN}(0,[1~ 1~ 1])$,
$\tilde{h}_{22}\sim\mathcal{CN}(0,[1~ 1~ 1])$,
$\tilde{h}_{33}\sim\mathcal{CN}(0,[1~ 1~ 1])$. Channel parameters 2: $\tilde{h}_{12}\sim \mathcal{CN}(0,[0.4~ 0.5~ 0.6])$,
$\tilde{h}_{13}\sim \mathcal{CN}(0,[0.5~ 0.5~ 0.3])$,
$\tilde{h}_{21}\sim \mathcal{CN}(0,[0.6~ 0.5~ 0.6])$,
$\tilde{h}_{31}\sim \mathcal{CN}(0,[0.7~ 0.5~ 0.4])$,
$\tilde{h}_{23}\sim \mathcal{CN}(0,[0.5~ 0.3~ 0.9])$,
$\tilde{h}_{32}\sim \mathcal{CN}(0,[0.4~ 0.5~ 0.6])$,
$\tilde{h}_{11}\sim \mathcal{CN}(0,[2~ 2~ 2])$,
$\tilde{h}_{22}\sim \mathcal{CN}(0,[1~ 1~ 1])$,
$\tilde{h}_{33}\sim \mathcal{CN}(0,[1~ 1~ 1])$.}
\label{sim fig3}
\end{figure}

On one hand, we evaluate the convergence performance in different channel states with the same ISR constraint, the same total power constraints and the same step-size. In the simulations, we set $N=3$, $N_1=N_2=N_3=[1~ 1~ 1]$, $\rho=0.2$, $P_1^{\max}=15$, $P_2^{\max}=5$, $P_3^{\max}=6$, and step-size $\delta=\eta=0.1$.
Fig. \ref{sim fig3} plots the convergence performance of Algorithm 2 and Algorithm 3 with different channel parameters. The PU and SUs are uniformly located in a square area of $10\times10$.
The channel gains are generated as $h_{i,j}=d_{i,j}^{-\alpha}\tilde{h}_{i,j}$, where $d_{i,j}$ represents the distance between the transmitter
of User $i$ and the receiver of User $j$, and $\alpha=2$ is the path loss\footnote{$\alpha=2$ corresponds to to free-space propagation.}.
It is observed that both Algorithm 2 and Algorithm 3 converge to the same results with the channel parameters 1 and channel parameters 2, respectively. Algorithm 2 converges since about the 50-th iteration, and Algorithm 3 converges since the 100-th iteration.
We should notice that there are 10 inner iterations in each iteration of Algorithm 2, then Algorithm 3 is more time-efficient. Moreover,
we can see that the rate performance with the channel parameters 2 is better. This can be explained as follows:
Comparing the channel parameters used in the simulations, there is stronger interference in the channel parameters 1. Then the performance with the channel parameters 2 will be better.
\par
\begin{figure}[]
\centering
\includegraphics[width=4in]{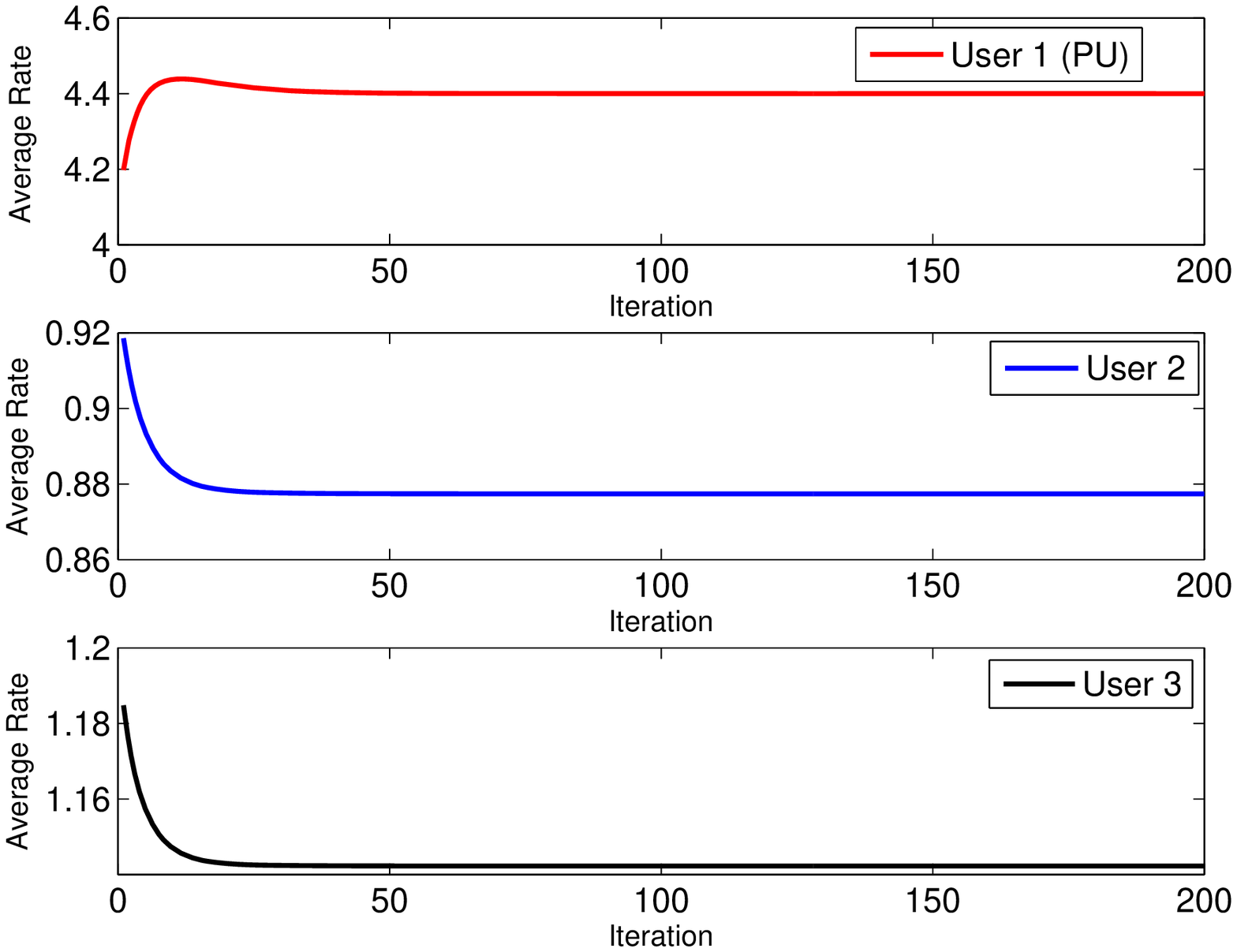}
\caption{Convergence performance of Algorithm 2 with  step-size $\eta=0.1$}
\label{sim fig51}
\end{figure}
\begin{figure}[]
\centering
\includegraphics[width=4in]{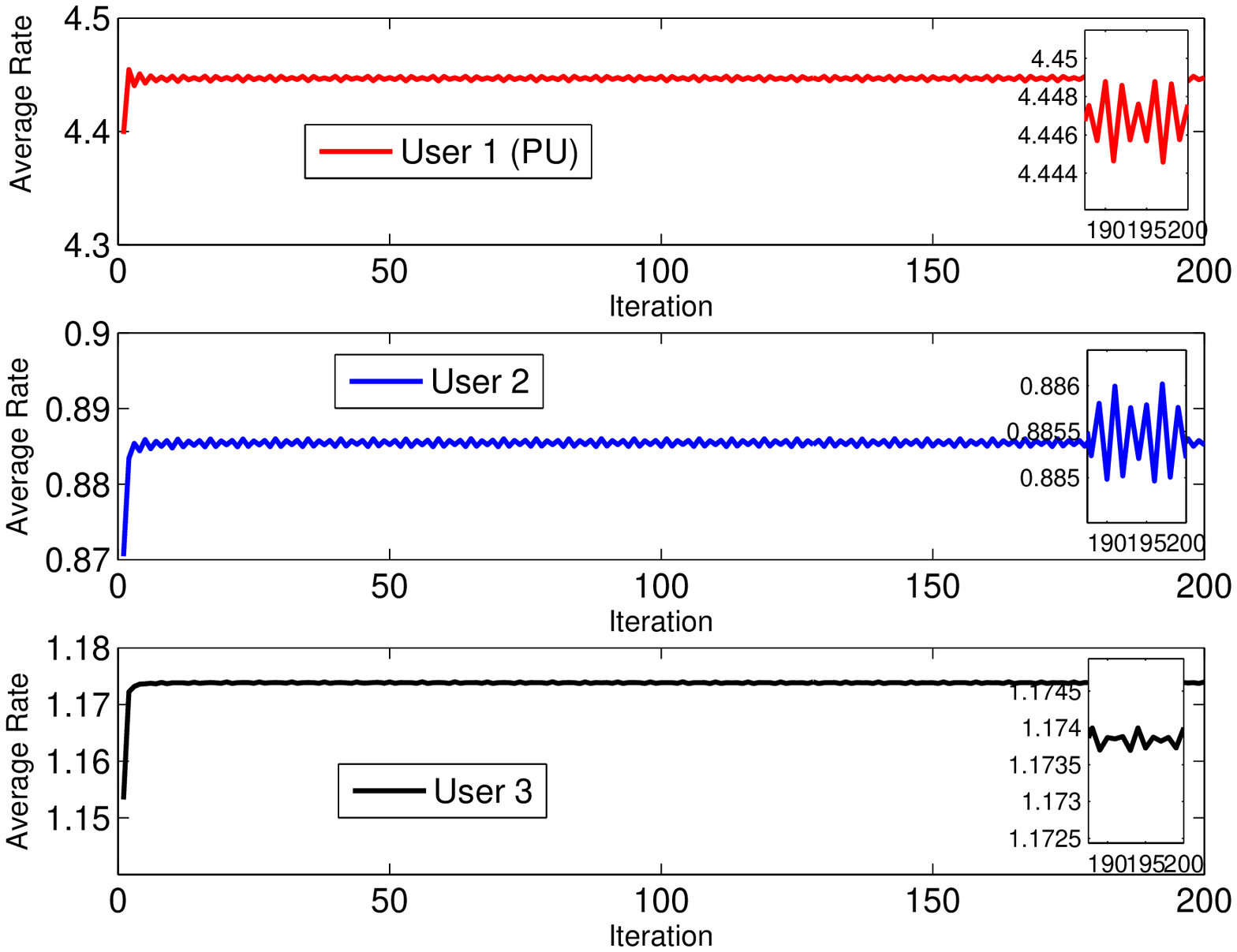}
\caption{Convergence performance of Algorithm 2 with step-size $\eta=0.9$}
\label{sim fig54}
\end{figure}
\begin{figure}[]
\centering
\includegraphics[width=4in]{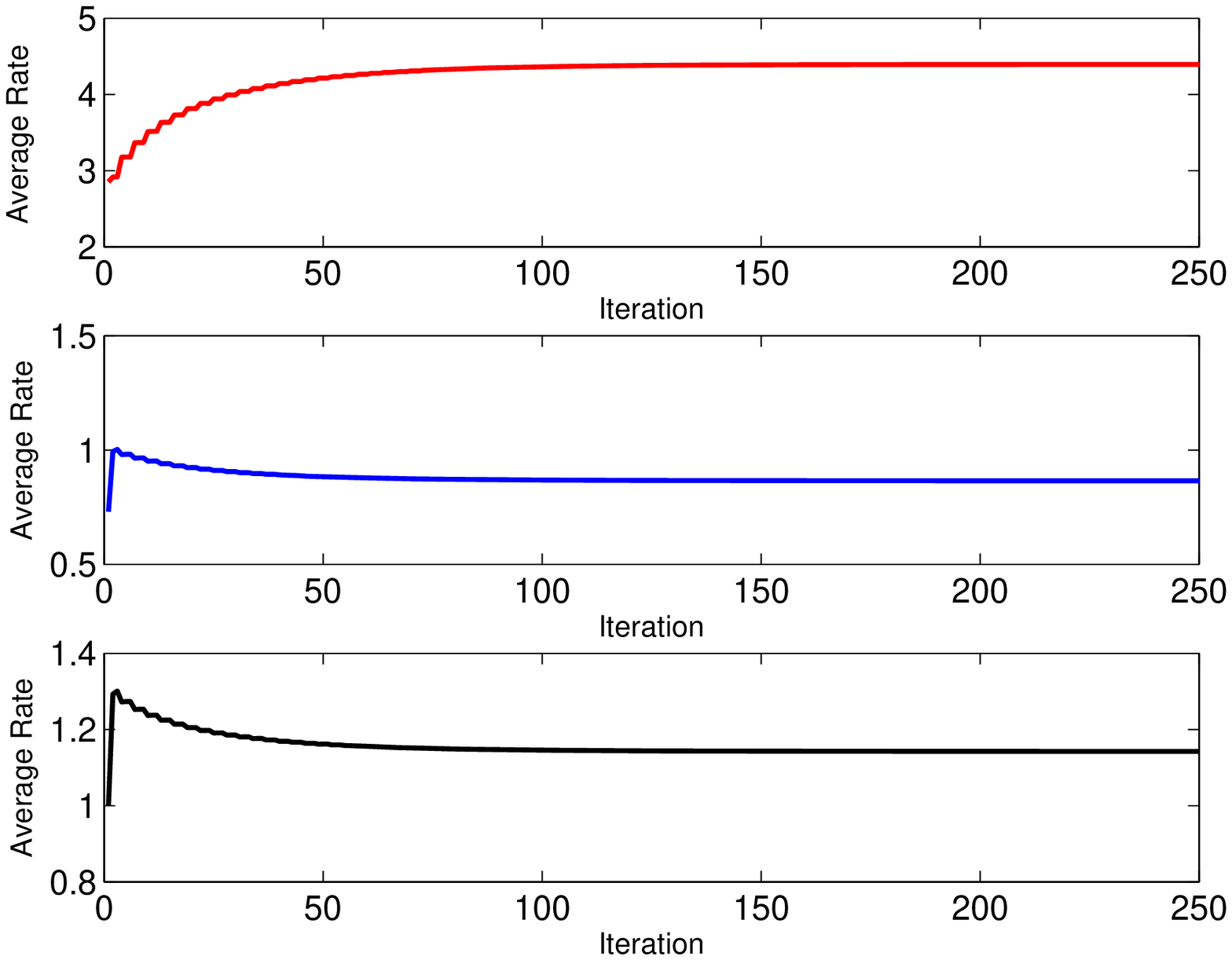}
\caption{Convergence performance of Algorithm 3 with step-size $\delta=0.1$}
\label{sim fig61}
\end{figure}
\begin{figure}[]
\centering
\includegraphics[width=4in]{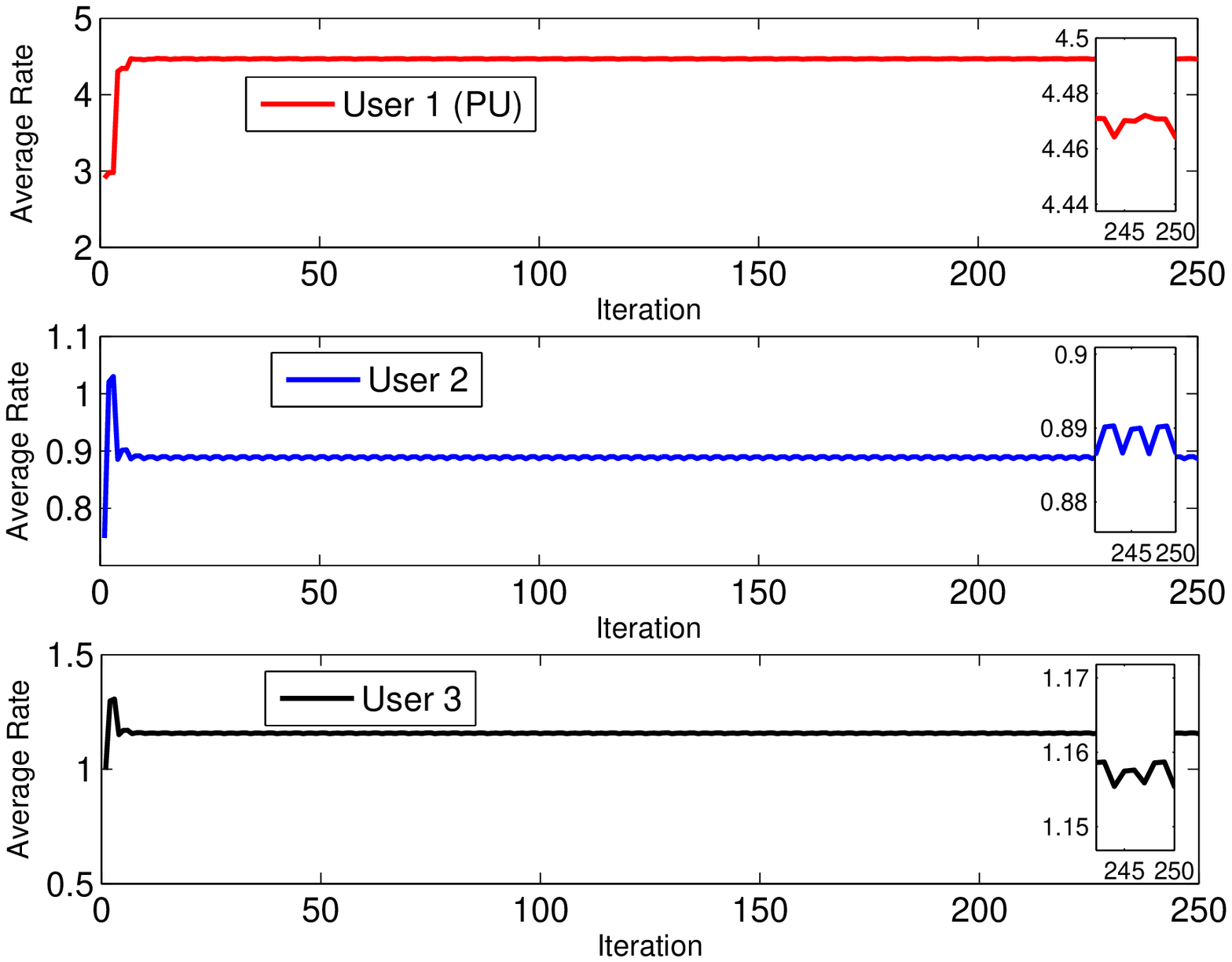}
\caption{Convergence performance of Algorithm 3 with step-size $\delta=0.9$}
\label{sim fig63}
\end{figure}

On the other hand, we evaluate the convergence performance in the same channel state with different step-sizes.
Parameters are chosen as follows:
$N=3$, $N_1=N_2=N_3=[1~ 1~ 1]$, $\rho=0.1$, $P_1^{\max}=25$, $P_2^{\max}=3$, $P_3^{\max}=4$,
$h_{12}\sim \mathcal{CN}(0,[0.4~ 0.5~ 0.6])$,
$h_{13}\sim \mathcal{CN}(0,[0.5~ 0.5~ 0.3])$,
$h_{21}\sim \mathcal{CN}(0,[0.6~ 0.5~ 0.6])$,
$h_{31}\sim \mathcal{CN}(0,[0.7~ 0.5~ 0.4])$,
$h_{23}\sim \mathcal{CN}(0,[0.5~ 0.5~ 0.5])$,
$h_{32}\sim \mathcal{CN}(0,[0.5~ 0.5~ 0.5])$,
$h_{11}\sim \mathcal{CN}(0,[1~ 1~ 1])$,
$h_{22}\sim \mathcal{CN}(0,[1~ 1~ 1])$,
$h_{33}\sim \mathcal{CN}(0,[1~ 1~ 1])$.
Fig. \ref{sim fig51} and Fig. \ref{sim fig54} demonstrate the convergence performance of Algorithm 2 with different step-sizes. We can observe that the algorithm converges with step-size $\eta=0.1$. However, when $\eta=0.9$, the algorithm does not converge, it oscillates. It can be interpreted by using Lemma \ref{convergence of A2}.
The upper bound for convergent step-size for all channel realizations lies between $0.1$ and $0.9$, i.e., $0.1<\min \xi<0.9$,
so when $\eta=0.1$, the condition in Lemma \ref{convergence of A2} can be satisfied, then the algorithm converges for all channel realizations and the average rate converges.
 When $\eta=0.9$, $\eta<\xi$ does not hold, the convergence cannot be guaranteed.
Fig. \ref{sim fig61} and Fig. \ref{sim fig63} illustrate the convergence performance of Algorithm 3 with different step-sizes. Similarly, we observe that the algorithm converges when the step-size is set to be $0.1$,
and it oscillates when the step-size equals to $0.9$.
\par
\begin{figure}[]
\centering
\includegraphics[width=4.5in]{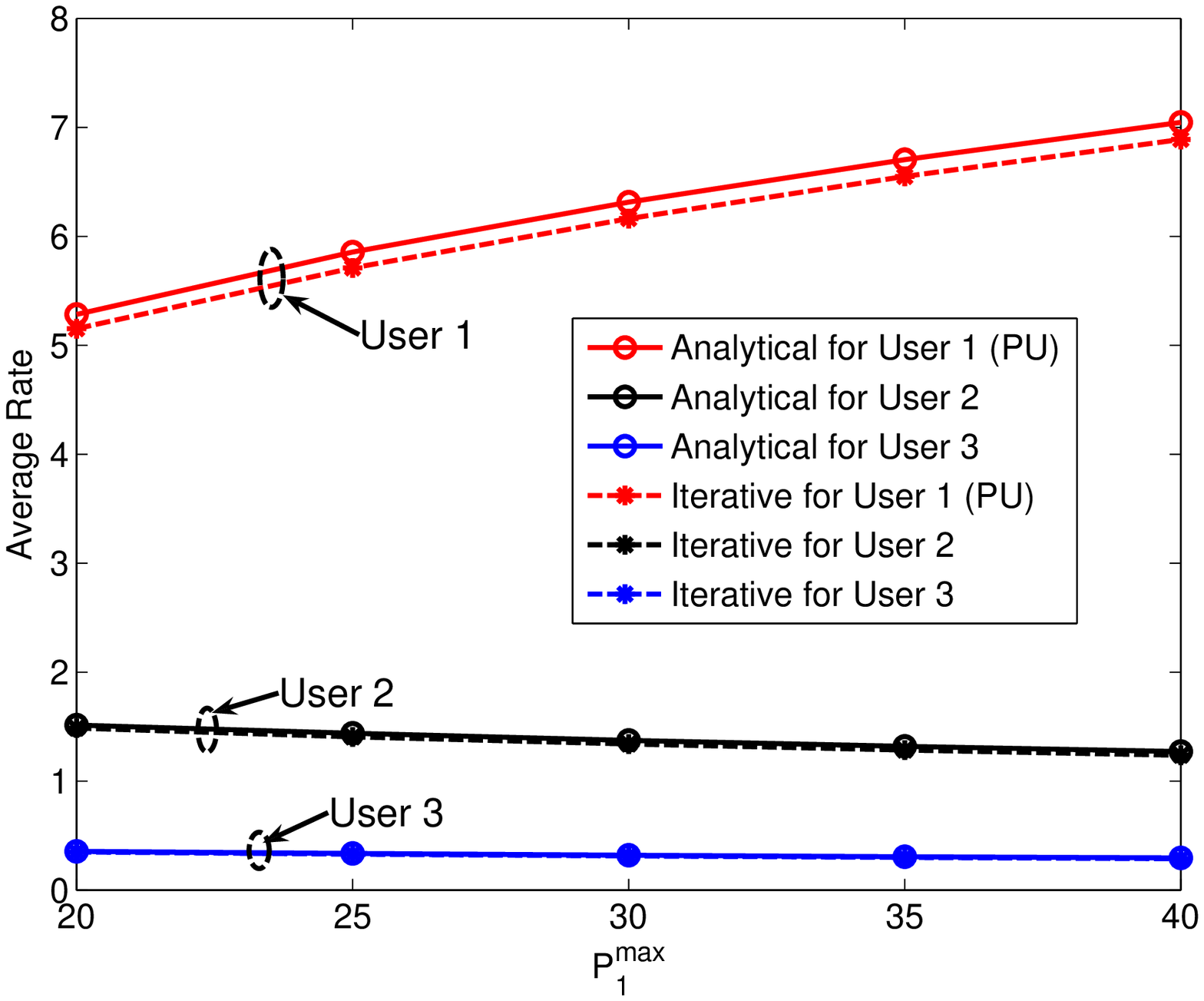}
\caption{The rate performance of the analytical power allocation and iterative power allocation in the perfectly symmetric channel with different $P_1^{\max}$. The other parameters are: $N=3$, $N_1=N_2=N_3=[0.5~ 0.5~ 0.5]$, $\rho=0.1$, $P_2^{\max}=5$, $P_3^{\max}=1$, $h_{11}\sim\mathcal{CN}(0,[1~ 1~ 1])$,
$h_{22}=h_{33}\sim \mathcal{CN}(0,[1~ 1~ 1])$,
$h_{12}=h_{13}\sim \mathcal{CN}(0,[\sqrt{0.2}~ \sqrt{0.3}~ \sqrt{0.4}])$,
$h_{21}\sim \mathcal{CN}(0,[0.3~ 0.6~ 0.5])$,
$h_{31}\sim \mathcal{CN}(0,[0.4~ 0.5~ 0.4])$,
$h_{23}=h_{32}=0.5\times h_{22}$ (i.e., $c=0.25$).
}
\label{sim fig7}
\end{figure}

In the perfectly symmetric channel, both analytical and iterative power allocation for PU and SU can be applied\footnote{Analytical method is applied when private information is available, and iterative method is used otherwise.}.
Fig. \ref{sim fig7} shows the rate performance of the analytical hierarchic power allocation and iterative power allocation for the PU and SUs with different power constraint for the PU, $P_1^{\max}$. We can observe that the rate performance of the PU decreases slightly in the iterative power allocation because of the unavailability of SUs' private information, but the rate performance of the SUs is almost the same as the analytical algorithm. This verifies effectiveness of the iterative power allocation.

%
%
%
%
%
%
\section{Extension to the multi-PU and multi-SU scenario}
When considering the multi-PU scenario, there are multiple leaders in the Stackelberg game,
they compete with each other to maximize their individual utility. Each PU considers not only the power allocation of other PUs, but also the rational reaction of SU network to the power allocation of the PU network. And we need to guarantee all PUs' ISR constraints. By minor adjustments, the proposed algorithms can be applied in the multi-PU and multi-SU scenario. In Algorithm 1, SU $i$ still measure the aggregated received interference, but the interference is generated by all PUs and other SUs in this scenario. In Algorithm 2 and Algorithm 3, the update of each PU's power can still utilize the former method. But the received interference should take other PUs' power allocation into consideration. In Algorithm 2, the convergence of PUs' power allocation should be achieved before the next iteration in multi-PU case. A renewed algorithm of Algorithm 2 for multi-PU is outlined as Algorithm 4.
\par
\begin{figure}[t]
\centering
\includegraphics[width=4.5in]{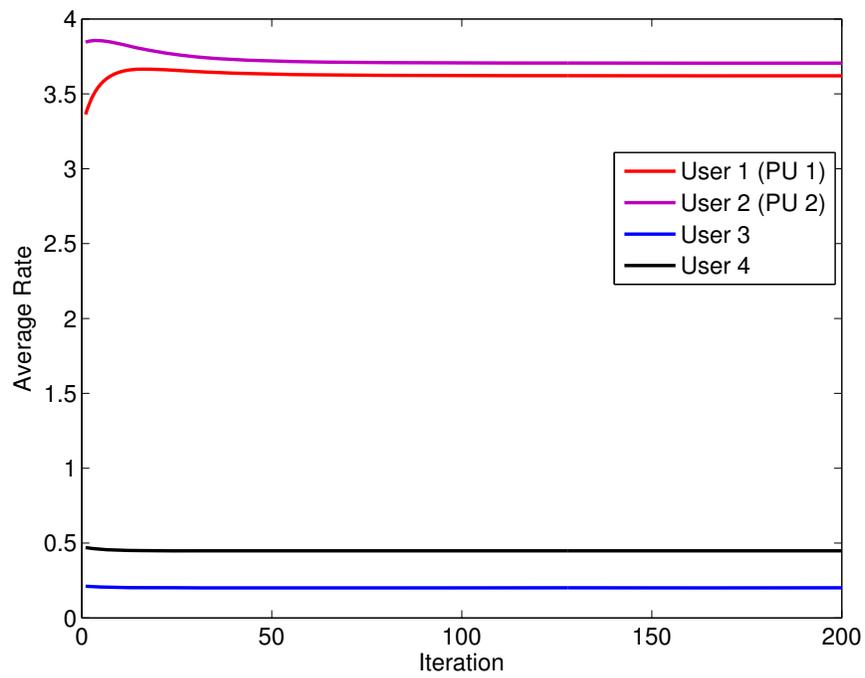}
\caption{Rate performance of the distributed iterative algorithm (Algorithm 4) for the multi-PU and multi-SU scenario}
\label{sim fig8}
\end{figure}

Fig. \ref{sim fig8} plots the rate performance when there are 2 PUs (user 1 and user 2) and 2 SUs (user 3 and user 4). In the simulation, the parameters are chosen as follows: $N=3$, $\rho=0.1$,
$P_1^{\max}=P_2^{\max}=15$, $P_3^{\max}=2$, $P_4^{\max}=6$,
$N_1=N_2=N_3=N_4=[1~ 1~ 1]$, $h_{12}\sim\mathcal{CN}(0,[0.5~ 0.2~ 0.1])$, $h_{13}\sim\mathcal{CN}(0,[0.5~ 0.5~ 0.3])$, $h_{14}\sim\mathcal{CN}(0,[0.4~ 0.5~ 0.6])$,
$h_{21}\sim\mathcal{CN}(0,[0.1~ 0.6~ 0.1])$, $h_{23}\sim\mathcal{CN}(0,[0.5~ 0.6~ 0.3])$, $h_{24}\sim\mathcal{CN}(0,[0.6~ 0.6~ 0.5])$,
$h_{31}\sim\mathcal{CN}(0,[0.3~ 0.7~ 0.2])$, $h_{32}\sim\mathcal{CN}(0,[0.2~ 0.1~ 0.5])$, $h_{34}\sim\mathcal{CN}(0,[0.6~ 0.8~ 0.6])$,
$h_{41}\sim\mathcal{CN}(0,[0.4~ 0.3~ 0.2])$, $h_{42}\sim\mathcal{CN}(0,[0.2~ 0.3~ 0.3])$, $h_{43}\sim\mathcal{CN}(0,[0.5~ 0.6~ 0.7])$,
$h_{11}\sim\mathcal{CN}(0,[1~ 1~ 1])$, $h_{22}\sim\mathcal{CN}(0,[1~ 1~ 1])$, $h_{33}\sim\mathcal{CN}(0,[0.5~ 0.5~ 0.5])$, $h_{44}\sim\mathcal{CN}(0,[0.5~ 0.5~ 0.5])$ and the step-size $\eta_i=0.001$ for $i \in \mathbb{P}$. The average rate is averaged over $10^5$ channel realizations.
From Fig. \ref{sim fig8}, we can see that the algorithm converges from the 60-th iteration.

\section{Conclusion}
We consider the power allocation for the PU network and SU network jointly by using the Stackelberg game to describe the hierarchy.
The PU network is considered as the leader, and the SU network acts as the follower.
We consider the ISR constraint to guarantee the primary service in the Stackelberg game. Based on the analysis of the Stackelberg game, the hierarchic power allocation algorithms are given. Analytical method is presented
when PU can obtain the information for SU. Once PU cannot obtain the information for SU, distributed iterative methods are proposed.

%
%
%
%
%
%
%
%

%
%
%


\ifCLASSOPTIONcaptionsoff
  \newpage
\fi



%
\begin{spacing}{1}

\end{spacing}

 \begin{table}[]
 \centering
 \begin{tabular}{lcl}
  \toprule
  \textbf{Algorithm 1: Iterative Distributed Algorithm} \\
  ~~~~~~~~~~~~~~~~~\textbf{for Obtaining NE}\\
  \midrule
 Step 1:  $k=0$, \\
  ~initialize feasible $\left\{\mathbf{P}_{i}(0)=\left(P_i^1(0),\cdots,P_i^N(0)\right)\right\}_{i \in \Omega}$.   \\
 Step 2: $P_i^f(k+1)=BR_i\left(P_1^f,P_{-i}^{f}(k)\right)$ \\
 ~for every $i \in \Omega$ and $f=1,\cdots,N$.\\
 Step 3: $k=k+1$, go to Step 2 until convergence. \\
  \bottomrule
 \end{tabular}
\end{table}

\begin{table}[]
 \centering
 \begin{tabular}{lcl}
  \toprule
   \textbf{Algorithm 2: Joint Iterative Distributed Power Allocation  }\\
   ~~~~~~~~~~~~~\textbf{Algorithm for PU and SUs (single-PU and multi-SU)}\\
  \midrule
 Step 1:  $n=0$, initialize  $\mathbf{P}_{1}(0)=\left(P_1^1(0),\cdots,P_1^N(0)\right)$. \\
 Step 2: Given $\mathbf{P}_{1}(n)$, the SUs allocate the NE power according    \\
         ~to (\ref{ne 2}) and (\ref{ne 3}) when the perfectly symmetric conditions can be      \\
         ~satisfied in the 2-SU scenario. Otherwise, the SUs apply       \\
         ~Algorithm 1 in the general scenario. Denote the allocated power \\
         ~for SUs as $\left\{\mathbf{P}_{i}(n)=\left(P_i^1(n),\cdots,P_i^N(n)\right)\right\}_{i \in \mathbb{S}}$.\\
 Step 3:  Update PU's power by using , $P_1^f(n+1)=$\\
 ~$(1-\eta)P_1^f(n)+\eta\left[\frac{I^f(n)}{\rho|h_{1,1}^f|^2}+\left(\lambda-\frac{I^f(n)+N_1^f}{|h_{11}|^2}\right)^+\right]$,\\
 ~where $\lambda$ is a constant to meet \\
 ~$\sum\nolimits_{f=1}^{N} \left[\frac{I^f(n)}{\rho|h_{1,1}^f|^2}+\left(\lambda-\frac{I^f(n)+N_1^f}{|h_{11}|^2}\right)^+\right] \le P_1^{\max}$, i.e., \\
 ~$\sum\nolimits_{f=1}^{N}\left(\lambda-\frac{I^f(n)+N_1^f}{|h_{11}^f|^2}\right)^+ \le P_1^{\max}-\sum\nolimits_{f=1}^{N}\frac{I^f(n)}{\rho|h_{1,1}^f|^2}$, \footnotemark  \\
  ~and $\eta \in (0,1)$ is a fixed step-size.\\
 Step 4: $n=n+1$, go to Step 2 until convergence. \\
  \bottomrule
 \end{tabular}
\end{table}
\footnotetext{$P_1^{\max}\ge \sum\nolimits_{f=1}^{N}\frac{I^f(n)}{\rho|h_{1,1}^f|^2}$ is assumed in this paper. }

\begin{table}[t]
 \centering
 \begin{tabular}{lcl}
  \toprule
   \textbf{Algorithm 3: Asynchronous Joint Iterative Distributed Power  }\\
   \textbf{Allocation Algorithm for PU and SUs (single-PU multi-SU)}\\
  \midrule
 Step 1:  $n=0$, $k=1$, initialize $\mathbf{P}_{1}(0)=\left(P_1^1(0),\cdots,P_1^N(0)\right)$ and \\
 $\left\{\mathbf{P}_{i}(0)=\left(P_i^1(0),\cdots,P_i^N(0)\right)\right\}_{i \in \mathbb{S}}$, $\mathbf{P}_{1}(0)$ and $\left\{\mathbf{P}_{i}(0)\}_{i \in \mathbb{S}}\right\}$ \\
  satisfy  their respective total power constraints and the\\
   ISR constraint.\\
 Step 2: Given $\mathbf{P}_{1}(n)$, the SUs update power allocation \\
 ~$\left\{\mathbf{P}_{i}(n+1)=\left(P_i^1(n+1),\cdots,P_i^N(n+1)\right)\right\}_{i \in \mathbb{S}}$ according to   \\
         (\ref{ne 2}) and (\ref{ne 3}) in the 2-SU scenario when the perfectly symmetric     \\
         conditions can be satisfied. Otherwise \\
         $P_i^f(n+1)=BR_i\left(P_{1}^{f}(n),P_{-i}^{f}(n)\right)$ \\
         for
         every $i \in \mathbb{S}$ and $f=1,\cdots,N$.  \\
 Step 3: Let $\{\tau_k\}_{k=1}^{\infty}$ be a subsequence of $\{n\}_{n=0}^{\infty}$ with \\
 $\tau_{k+1}-\tau_k<\infty$
  for finite $k$. \\
  The PU updates its power asynchronously by \\
$P_1^f(n+1)=$\\
 $\left\{ \begin{array}{l}
 (1-\delta)P_1^f(n)+\delta \left[\frac{I^f(n+1)}{\rho|h_{1,1}^f|^2}+\left(\lambda-\frac{I^f(n+1)+N_1^f}{|h_{11}|^2}\right)^+\right],\\
 ~~ n=\tau_k ; \\
 P_1^f(n), \mathrm{otherwise}.
\end{array}
\right.
$ \\
 $f=1,2,\cdots,N$, where
 $\delta \in (0,1)$ is the fixed step size. \\
 If $n=\tau_k$, $k=k+1$.\\
 Step 4: $n=n+1$, go to Step 2 until convergence or $n=N_{max}$. \\
  \bottomrule
 \end{tabular}
\end{table}

\begin{table}[!t]
 \centering
 \begin{tabular}{lcl}
  \toprule
   \textbf{Algorithm 4: Joint Iterative Distributed Power Allocation  }\\
   ~~~~~~\textbf{Algorithm for PUs and SUs (multi-PU and multi-SU) }\\
  \midrule
 Step 1:  \\
 ~~$n=0$, initialize  $\mathbf{P}_{i}(0)=\left(P_1^1(0),\cdots,P_1^N(0)\right)$, $i \in \mathbb{P}$. \\
 Step 2: \\
 ~~Given $\{\mathbf{P}_{i}(n)\}_{i \in \mathbb{P}}$, the SUs allocate the NE power according to   \\
         ~~(\ref{ne 2}) and (\ref{ne 3}) when the perfectly symmetric conditions can be      \\
         ~~satisfied in the 2-SU scenario. Otherwise, the SUs apply  \\
         ~~Algorithm 1 in the general scenario (Observe that $P_1^f|h_{1,i}^f|^2$  \\
          ~~should be replaced by $\sum\nolimits_{l \in \mathbb{P}}P_l^{f}(n)|h_{l,i}^f|^2$).\\
         ~~Denote the allocated power for SUs as \\
         ~~$\left\{\mathbf{P}_{i}(n)=\left(P_i^1(n),\cdots,P_i^N(n)\right)\right\}_{i \in \mathbb{S}}$.\\
 Step 3:\\
    ~~Sub-step 3.1:
 $k=0$, $\mathbf{P}_{i}(k)=\mathbf{P}_{i}(n)$ for all $i \in \mathbb{P}$.\\
 ~~Sub-step 3.2: For every $i \in \mathbb{P}$, PU $i$ updates its power by using \\
 ~~$P_i^f(k+1)$\\
 ~~~~~~$=(1-\eta_i)P_i^f(k)+\eta_i\left[\frac{I_i^f(k)}{\rho|h_{i,i}^f|^2}+\left(\lambda_i-\frac{I_i^f(k)+N_i^f}{|h_{ii}^f|^2}\right)^+\right]$,\\
 ~~where $I_i^f(k)=\sum\nolimits_{l \ne i \in \mathbb{P}}P_l^{f}(k)|h_{l,i}^f|^2+\sum\nolimits_{j \in \mathbb{S}}P_j^{f}(n)|h_{j,i}^f|^2$ is   \\
 ~~the total received interference, $\lambda_i$ is a constant to meet \\
 ~~$\sum\nolimits_{f=1}^{N} \left[\frac{I_i^f(k)}{\rho|h_{i,i}^f|^2}+\left(\lambda_i-\frac{I_i^f(k)+N_i^f}{|h_{ii}|^2}\right)^+\right] \le P_i^{\max}$, i.e., \\
 ~~$\sum\nolimits_{f=1}^{N}\left(\lambda_i-\frac{I_i^f(k)+N_i^f}{|h_{ii}^f|^2}\right)^+ \le P_i^{\max}-\sum\nolimits_{f=1}^{N}\frac{I_i^f(k)}{\rho|h_{i,i}^f|^2}$,  \\
  ~~and $\eta_i \in (0,1)$ is a fixed step-size.\\
  ~~Sub-step 3.3: $k=k+1$, go to Sub-step 3.2 until convergence.\\
  ~~Sub-step 3.4: $\mathbf{P}_{i}(n+1)=\mathbf{P}_{i}(k)$ for $i \in \mathbb{P}$.\\
 Step 4: \\
 ~~$n=n+1$, go to Step 2 until convergence or $n=N_{max}$. \\
  \bottomrule
 \end{tabular}
\end{table}

\end{document}